\documentclass[12pt,preprint]{aastex}

\slugcomment{Accepted for publication in ApJ - 06 June 2006}

\shorttitle{HD 181327 Debris Ring}
\shortauthors{Schneider, et al.}

\begin{document}

\title{Discovery of an 86 AU Radius Debris Ring Around HD 181327}

\author{Glenn Schneider\altaffilmark{1}, 
Murray D. Silverstone\altaffilmark{1}, 
Dean C. Hines\altaffilmark{2}, 
Jean-Charles Augereau\altaffilmark{3}, 
Christophe Pinte\altaffilmark{3}, 
Fran\c cois M\'enard\altaffilmark{3}, 
John Krist\altaffilmark{4}, 
Mark Clampin\altaffilmark{5},
Carol Grady\altaffilmark{6}, 
David Golimowski\altaffilmark{7}, 
David Ardila\altaffilmark{8},
Thomas Henning\altaffilmark{9},  
Sebastian Wolf\altaffilmark{9}, 
Jens Rodmann\altaffilmark{9}}

\altaffiltext{1}{Steward Observatory, The
University of Arizona, 933 N. Cherry Ave., Tucson, AZ 85721 USA}

\altaffiltext{2}{Space Science Institute, 4750 Walnut Street,
Suite 205 Boulder, CO 80301 USA}

\altaffiltext{3}{Laboratoire d'Astrophysique de Grenoble, B.P. 53, 38041 Grenoble Cedex 9, France}

\altaffiltext{4}{Jet Propulsion Laboratory, 4800 Oak Grove Dr., M/S 183-900 Pasadena, CA 91109}

\altaffiltext{5}{NASA/Goddard Space Flight Center, Code 680 Greenbelt, MD 20771}

\altaffiltext{6}{Eureka Scientific,10813 Graeloch Rd., Laurel, MD 20723-1124}

\altaffiltext{7}{Johns Hopkins University, Baltimore, MD 20218}

\altaffiltext{8}{Spitzer Science Center, 1200 E. California Blvd., Pasadena, CA 91125}

\altaffiltext{9}{Max-Planck-Institut fur  Astronomie, Konigstuhl 17,D-69117 Heidelberg}

\begin{abstract}

{\it HST}/NICMOS PSF-subtracted coronagraphic observations of HD~181327 have revealed the presence of a ring-like disk of circumstellar debris seen in 1.1 $\mu$m light scattered by the disk grains, surrounded by a diffuse outer region of lower surface brightness. The annular disk appears to be inclined by 31\fdg7 $\pm$ 1\fdg6 from face on with the disk major axis PA at 107$\degr$ $\pm$ 2$\degr$. The total 1.1 $\mu$m flux density of the light scattered by the disk (at  1\farcs2 $<$ {\it r} $<$ 5\farcs0) of  9.6 mJy  $\pm$ 0.8 mJy  is  0.17\% $\pm$ 0.015\% of the starlight. Seventy percent of the light from the scattering grains appears to be confined in a 36 AU wide annulus centered on the peak of the radial surface brightness (SB) profile 86.3 AU $\pm$ 3.9 AU from the star, well beyond the characteristic radius of thermal emission estimated from {\it IRAS} and {\it Spitzer} flux densities assuming blackbody grains ($\approx$ 22 AU). The 1.1 $\mu$m  light scattered by the ring: (a) appears bilaterally symmetric,  (b) exhibits directionally preferential scattering well represented by a Henyey-Greenstein scattering phase function with g$_{HG}$ = 0.30 $\pm$ 0.03, and (c) has a median SB (over all azimuth angles) at the 86.3 AU radius of peak SB of 1.00  $\pm$ 0.07 mJy arcsec$^{-2}$.  No photocentric offset is seen in the ring relative to the position of the central star. A low surface brightness diffuse halo is seen in the NICMOS image to a distance of $\sim$ 4$\arcsec$. Deeper 0.6  $\mu$m {\it HST}/ACS PSF-subtracted coronagraphic observations reveal a faint  (V $\approx$ 21.5 mag arcsec$^{-2}$) outer nebulosity from 4$\arcsec < r < 9\arcsec$, asymmetrically brighter to the North of the star.  We discuss models of the disk and properties of its grains, from which we infer a maximum vertical scale height of 4--8 AU at the 87.6 AU radius of maximum surface density, and a total maximum dust mass of collisionally replenished grains with minimum grain sizes of $\approx 1~\mu$m of $\approx$ 4 M$_{Moon}$.

\end{abstract}





\section{INTRODUCTION}

Fifteen years elapsed between the acquisition of the prototypical scattered-light image of a debris disk around a young main-sequence star ($\beta$ Pictoris; Smith \& Terrille 1984), offering strong support to the conjecture of the ``Vega phenomenon'' (far-IR emission in excess of a stellar photosphere from orbiting thermally emissive dust \citep{BAC93} as suggested from earlier {\it IRAS} observations \citep{aumann84}), and the second example of a dusty circumstellar disk -- in the form of a ring -- seen by starlight illuminating its constituent grains (HR 4796A; Schneider et al. 1999).  Advances in space- and ground- based high contrast imaging capabilities have given rise to increasing, though still very small, numbers of scattered-light images of circumstellar debris systems. The very recent discoveries of a high surface brightness $\beta$ Pictoris-like nearly edge-on disk around the A-star HD 32297 (Schneider, Silverstone \& Hines 2005 [=SSH05]), and the very low surface brightness 140 AU ring around an A-star much older than $\beta$ Pictoris, i.e., Fomalhaut \citep{KAL05} speak  to the diversity in disk morphologies and scattering properties. While ring-like structures are obscured in scattered-light images of nearly edge-on disks, their intrinsic morphologies may be inferred by inversion of their radial surface brightness profiles (egs.: \cite{buit}, \cite{pantin96}, \cite {aug01} for $\beta$ Pictoris; and models presented by \cite{krist05} and \cite{metch05} for AU Microscopii).
 We now report the discovery of a ring-like disk of circumstellar debris seen about the star HD~181327. This new addition to the growing menagerie of circumstellar debris disks is, in many respects, similar to the HR~4796A star:disk system.  The  recent imagery of a broader ring about the older solar analog, HD~107146 \citep{ardila04}, suggests that such structures are a common (if not prevalent) outcome of the planet formation process.  

HD 181327 (HIP 95270) is an F5/F6V star ({\it T}$_{eff}$ $\approx$ 6450 K; \cite{nord04}, log($g$) = 4.5) at 50.6 pc $\pm$ 2.1 pc \citep{peri97} with a Vega-like far-IR excess noted by Backman \& Paresce (1993).  HD 181327  was identified as a main sequence debris-disk candidate by Manning \& Barlow (1998)  with an {\it L}$_{ir}$/{\it L}$_{*}$ = 0.2\%.  Kalas et al (2002) favored the Vega phenomenon interpretation of its far-IR excess based upon its youth and its gas depleted environment. The star's youth was first suggested from its potential  membership in the $\sim$ 40 Myr Tucanae association \citep{ZUK00}, but later identified as a member of the younger ($\sim$ 12 Myr) $\beta$~Pictoris moving group (Zuckerman \& Song 2004; Mamajek et al. 2004). 

The conjecture of a stellar age of $\sim$ 10$^{7}$ yr is further supported by two independent observations. First, HD~181327 appears to exhibit common proper motion  with the 7$\arcmin$ (0.1 pc at 50 pc) distant HR 7329, also identified as a member of the $\beta$ Pictoris moving group by \cite{ZUK01}. Both stars have statistically identical Hipparcos parallaxes and (long baseline) Tycho 2 proper motions. HR 7329 is a ZAMS A-star possessing a close substellar companion\footnote{The total mass of HD181327 + HR7329A/B is $\sim$ 3.8 M$_{Sun}$, and stellar binaries with separations $\sim$ 0.1 pc 
are not unheard of with similar systemic masses.  Hence, in light of recent discussion on brown dwarf formation and binary populations 
(egs., see  \cite{close03} fig 15 and \cite{burgasser05}) one might conjecturally consider HD181327 + HR7329A/B to be a triple system.} 
with an isochronal age of $\sim$ 10--30 Myr \citep{Lowrance2000}. Second, HD 181327's Hipparcos derived B-V color index of +0.480 is completely consistent with an unreddeded F5--F6 star\footnote{In the Hipparcos catalog, the F5 and F6 stars within 75 pc with parallaxes of S/N $\geq$ 8 have median B-V color indices of +0.464 and+0.496, respectively. Hence, there appears to be no evidence for reddening for HD 181327 (unsurprising given its distance and environment).}. This, then, is in agreement with HD181327's e(b-y) = 0.002 mag \citep{nord04}, and the stellar luminosity  is then 3.1 $\pm$ 0.3 L$_{sun}$ from which the stellar age estimated using evolutionary tracks by \citet{dm97}, \citet{ps01}, and \citet{ba98} is consistent with $\sim$ 20 Myr.

Assuming blackbody grains, \cite{ZUK04} suggested a thermal equilibrium temperature of 65 K for the IR emissive dust at a characteristic orbital radius 35 AU from HD 181327. 
\citet{NEU04}, using  {\it NNT} + SHARP-I \& SofI, and \citet{CHA04}, with {\it ESO} 3.6 m AO + ADONIS/SHARP-II, conducted searches for a companion candidate to HD 181327, but found none. 

\section{NICMOS OBSERVATIONS \& DATA PROCESSING}

During {\it HST} cycle 13 we conducted a NICMOS coronagraphic imaging survey ({\it HST} GO/10177; PI: Schneider) of 26, $\sim$ 10 Myr and older, main sequence stars with strong far-IR excess emission, including HD~181327, to identify and provide spatially resolved scattered-light images of previously suspected, but unimaged, dusty circumstellar disks (SSH05).   HD~181327 was observed at two field orientations in a single spacecraft orbit on 2005 May 02 UT using {\it HST}/NICMOS camera 2 (scale:  $\sim$ 75.8 mas pixel$^{-1}$) with F110W ($\lambda$$_{eff}$ = 1.104 $\mu$m, FWHM = 0.5915 $\mu$m) filter direct and coronagraphic imaging, and supporting F180M ($\lambda$$_{eff}$ = 1.7968 $\mu$m, FWHM = 0.0684 $\mu$m) target acquisition exposures.  The specifics of the observations, and those of the PSF reference stars from which our PSF-subtracted imagery  we discuss in \S 2.1, are given in Table 1.  Details of the observing strategy, data reduction and calibration methodologies, and image combination processes which we apply to all debris disk candidate (and reference PSF) star observations in our {\it HST} GO/10177 survey are discussed by SSH05 and to which we refer the reader for details.

\subsection{NICMOS PSF-Subtractions}
To further reduce the level of the residual instrumentally diffracted and scattered circumstellar light, beyond the improvement in image contrast afforded by the coronagraph itself, we subtract astrometrically registered, flux scaled images of coronagraphic reference PSFs of non-disk stars of similar spectral energy distributions from each of the two (one at each field orientation) calibrated combined images of HD 181327. We follow the precepts of the two-orientation coronagraphic PSF-subtraction process described by SSH05. We selected ten candidate PSF stars with near-IR colors similar to HD 181327. Two of these were specifically programmed very bright PSF calibration stars with no {\it IRAS} detected excess emission, HR 4748 (J=5.52 mag) and HIP 7345 (J=5.49 mag), each observed at a single field orientation using a full spacecraft orbit and yielding very high SNR. Eight other PSF template candidates were selected from GO 10177 survey targets that exhibited no evidence of scattered-light excesses from circumstellar dust.  Each of these eight stars were observed at two field orientations (thus providing two PSF templates each), but were much less deeply exposed than our PSF calibration targets due to their lesser brightness and the detailed nature of the observing/integration time strategy for these targets.   All 36 HD 181327-minus-PSF template subtractions revealed an elliptical ring-like disk that rotated in the celestial frame, following the 29\fdg9 difference in target orientation between the two HD 181327 observational visits (e.g., Fig 1A--D). PSF-minus-PSF subtractions exhibited no such structures (e.g., Fig 1E) while in all cases rotationally invariant features due to instrumental scattering and diffraction remained fixed in the frame of the detector. 

The four images resulting from subtracting the HR 4748 and HIP 4375 high SNR reference PSFs from each of the two images of HD 181327 (differing in orientation 29\fdg9) were of higher photometric efficacy than those obtained using the serendipitous ``ensemble'' of PSFs derived from the GO 10177 survey targets and, generally, were more devoid of azimuthally anisotropic high and mid spatial frequency artifacts arising from aliasing in imperfect PSF-subtractions. The HD 181327 debris disk image, which we discuss in this paper, was created using these four PSF-subtracted images.  The four images were median combined with equal weight after alignment, rotation to a common orientation, and masking regions in individual images affected by diffraction spikes from the HST secondary mirror support structure (see Figure 1F).  The final NICMOS image, shown in Figure 1F (and 2A), and discussed in \S\S~5 \& 6, has a total integration time of 1536 s in most pixels.

\section{ACS OBSERVATIONS}
HD 181327 was observed earlier in {\it HST}/GTO  program 9987 (PI: Ford) on 2003 November 02 UT using the {\it HST} Advanced Camera for Surveys/High Resolution Camera (ACS/HRC; scale:  $\sim$ 25 mas pixel$^{-1}$) in direct
and coronagraphic imaging modes.  Two 5 s exposures were taken without the
coronagraph in  the F606W (``wide'' V-band) filter.  After a 0.5 s F502N acquisition exposure,
the star was placed behind the 0\farcs 9 radius coronagraphic occulting spot
and two 1235 s F606W exposures were taken.  The ACS images were
calibrated by the {\it HST} OPUS\footnote{http://www.stsci.edu/resources/software\_hardware/opus/} pipeline, and the two coronagraphic
 exposures were combined with cosmic ray rejection.

\subsection{ACS Coronagraphy \& PSF Subtraction}
By suppressing the diffraction pattern of the occulted star, the ACS coronagraph
reduces the surface brightness of the wings of the stellar PSF by about a
factor of 7 -- 10.  As with NICMOS, an additional gain in image contrast can be 
achieved by subtracting the underlying stellar PSF using a coronagraphic image of another occulted star of similar color reducing the remaining ``background''
by factors of  10 -- 100.  Additional information on ACS coronagraphic performance
is presented in  \S~6.2 of \cite{gonz}. 

In the case of HD~181327, 
observations of the F5V star HD~191089 were used to create a template PSF image
for subtraction.  HD~191089 was imaged in a single spacecraft orbit immediately following
the observations of HD~181327, and a very similar imaging and data processing
strategy was employed.  Two 6 s direct  and two 1005 s coronagraphic (r = 0\farcs9 occulting spot)
exposures  were obtained with the F606W filter. The HD~191089 PSF template 
was scaled to the flux density of HD~181327\footnote{To establish the flux density scaling, for both stars the number of 
saturated pixels in the PSF core was multiplied by 165000 e$^{-}$ pixel$^{-1}$ 
(full well depth of an HRC CCD pixel) and added to the flux density
measured within a 1$\arcsec$ radius aperture. Charge in excess of a pixel full-well
is transferred, without significant inefficiency, to neighboring non-saturated pixels.  For additional information see $\S\S$ 4.2.6 and 4.3.1 of 
\cite{gonz}, \cite{gill04} and \cite{bohlin02}.}.  After finite-aperture photometric and geometrical
distortion corrections, the target-to-PSF scaling obtained agreed within 2\% of expected
based upon the V-band magnitudes of the stars.   The
subtraction of HD~181327 was optimized by iteratively shifting and subtracting
the flux-scaled HD~191089 image until residuals induced by mis-registration, beyond 2$\arcsec$
from the occulted star, were  minimized and spatially symmetric.  This process 
co-registers the stars to $\sim$ 0.05 HRC pixels (1.25 mas).  The final subtracted image (shown in figure 3 at three different image scales and display stretches) was corrected for the well-characterized geometric distortion in the HRC camera.

\subsection{ACS Image Data  \& Caveats}

Near the coronagraphic obscuration, a residual halo of starlight unsuppressed  by the ACS coronagraph arises from 
instrumental scattering  (seen as radial streaks) and diffraction by the occulter itself 
(seen as  concentric rings centered on the occulter).  The scattered-light streaks  are relatively insensitive to star-to-occulter
positioning, but the ring-like diffracted light distribution is sensitive to the
location of the star within the footprint of the occulter. The ACS PSF-subtracted image of 
HD~181327-minus-HD~191089
(after flux-scaling and image registration)  shows instrumentally-induced 
residuals beyond the edge of the occulting spot.  The residuals are  dominated by
diffraction rings arising, predominantly,  from  differences in the positioning of the 
two stars behind the occulter.   The registration of the HD~181327 and HD~191089 images required a larger than typical 
translation (30.1 mas) of the reference PSF indicative of a target centration error in the acquisition 
process (assuming the position of the occulting spot remained stable over the short period of the two-orbit 
sequential observations of both stars).
The position error was much greater along the image horizontal (w.r.t. the detector), so the PSF-subtraction residuals are also larger along that direction.
Due to the asymmetry of the PSF subtraction-residuals near the edge of the spot:
({\it a})  the HD~181327 disk can be detected reliably  from 1$\farcs 4 < r <  2\farcs0$ from the star 
only within  $\sim  \pm~10\degr$ arcs centered on the  image-Y axis on opposite sides of the star (figure 3A);
({\it b}) from 2$\arcsec < r < 4\arcsec$ a diffuse halo is seen 
(figure 3B), though its near circular-symmetry and ring-like substructures might implicate a residual 
contribution of instrumental origin (from imperfect PSF-subtraction) that is somewhat difficult to quantify. 

Further from the coronagraphic occulting spot, from 4$\arcsec < r <$  9$ \arcsec$ (well beyond the distance leading to PSF-subtraction artifacts discussed in ({\it a}) and ({\it b})),
within $\pm$ 45$\degr$ of North a 
faint nebulosity with a mean surface brightness of V $\approx$ 21.5 mag
arcsec$^{-2}$ is seen (figure 3C).

\section{{\it Spitzer}/MIPS and {\it IRAS} PHOTOMETRY}
 
New 24, 70, and 160 $\mu$m imaging photometric observations of HD 181327 were obtained with the MIPS instrument on the {\it Spitzer} Space Telescope on 2004 April 06 in GTO program 72 (F. Low, PI). We use the photometric measures derived from these observations, along with  {\it IRAS} Faint Source Catalog (FSC) 25, 60, and 100 $\mu$m photometry, to better define the thermal spectral energy distribution (SED) of the HD 181327 disk (See Fig. 4 and Table 2).  With these data we find a slightly larger thermal IR excess, {\it L}$_{ir}$/{\it L}$_{*}$ = 0.25\%, than Manning \& Barlow (1998), and a somewhat warmer blackbody equilibrium temperature for the disk grains than the 65 K suggested by \cite{ZUK04}. A least-squares blackbody fit to the 24 $\mu$m and longer flux densities indicates an equilibrium temperature of 77 K  $\pm$ 5 K for the thermally emissive  grains (after subtracting off the small underlying contribution from the stellar photosphere).  Following \cite{ZUK04}, the warmer grain temperature would also  suggest -- for grains emitting as blackbodies -- a smaller characteristic orbital radius of $\approx$ 22 AU. The 12 $\mu$m emission detected by {\it IRAS} is not in excess of the stellar photospheric level (see Fig 4) as confirmed from a recent {\it Spitzer}/IRS spectrum of HD 181327 \citep{chen}.

\section{THE HD 181327 RING-LIKE DISK}

The NICMOS 1.1 $\mu$m PSF-subtracted image of the HD~181327 disk probes closer to star than the ACS 0.6 $\mu$m imagery at all azimuth angles ($\theta$; with $\theta$ = 0$\degr$ at the position angle closest to the line-of-sight toward the Earth; see \S 5.1).  Because the effective inner radius of the ACS image was beyond the peak of the radial surface brightness (SB) distribution at most azimuth angles (as revealed by the NICMOS image), we determined the systemic geometrical and over-all photometric properties of the HD~181327 disk from the NICMOS image alone. 

The HD 181327 debris disk seen in the NICMOS 1.1 $\mu$m image is in the form of an inclined ring (figure 2A). The radial SB of the ring declines monotonically from the (deprojected) radius of its peak brightness, {\it R$_{peak}$}, both  inward (appearing cleared toward the center) and outward (appearing bounded or truncated) from {\it R$_{peak}$}.

To establish the geometry of the distribution of debris-scattered light in the HD~181327 ring, we fit  image isophotes to nested ellipses and simultaneously solved for the ring axial lengths, major axis orientation, and photocenter, by iterative $\chi^{2}$ minimization to the NICMOS image. The location of the star (occulted by the coronagraph) was determined using acquisition offset slew vectors downlinked in the engineering telemetry applied to star's position measured from the well-exposed and critically-sampled, pre-slew target acquisition images with a precision of $\lesssim$ 1/20 pixel ($\sim$ 4 mas; 0.2 AU at 50 pc).

\subsection{Geometry of the Ring}

{\it RING GEOMETRY.}  Assuming intrinsic circular symmetry, we found the ring inclination, {\it i} = 31\fdg7 $\pm$ 1\fdg6 from a face on viewing geometry, with the disk major axis oriented at a PA of 107$\degr$ $\pm$ 2$\degr$ east of north. The best elliptical fits find the radial peak in surface brightness at a semimajor axis length of 22.5 $\pm$ 0.3 pixels (1\farcs706 $\pm$ 0\farcs023).  Given HD 181327's  annual parallax of 19.77 $\pm$ 0.81 mas, we find the radius of the debris ring, {\it R$_{peak}$} = 86.3 $\pm$  3.7 AU.

{\it BILATERAL SYMMETRY.}  Figure 5 shows the disk deprojected from the plane of the sky with the major axis on the image horizontal and the location of the occulted star marked.
The HD 181327 ring surface brightness appears to be bilaterally symmetric, and the disk surface brightness peaks, radially, at the same distance at all azimuths. The absence of any significant departure from circular symmetry is illustrated in a low-pass filtered image (Figure 5B) and the corresponding  isophotal map of the deprojected disk  (Figure 5C). The ring appears to be centered on the star to within a measurement uncertainty of $\sim$ 0.25 pixels (1 AU at 50 pc).   

\subsection{Photometry of the Disk}

{\it 1.1 $\mu$m FLUX DENSITY.} Because of residual ``noise'' in the PSF subtraction close to the star, there is little useful information at {\it r}  $< $ 1\farcs2 from the star.  The flux density from the ring dominates over the subtraction residuals at larger radii.  We measured the total 1.1  $\mu$m flux density of the disk at 1\farcs2 $< $ {\it r} $< $ 5\farcs0 as 9.6 $\pm$ 0.8 mJy. The main contributor to the $\sim$~8\% measurement error arises from imperfect removal of the underlying stellar PSF by under- or over- subtraction due to uncertainties in flux scaling.  We assess the magnitude of this error from null  PSF-PSF subtractions of the reference stars,  scaled to the flux density of HD 181327 (e.g., see Schneider et al 2001 and SSH05). There may be more dust-scattered light interior to this radius. However, as will be shown, the radial surface brightness of the disk declines steeply inward from the brightest part of the ring at 1\farcs71 (deprojected). The small area interior to 1\farcs2 is likely to only weakly  contribute to the total light-scattered disk flux density.

{\it 1.1 $\mu$m SCATTERING FRACTION.} Our direct (non-coronagraphic) images of HD 181327, were  saturated in their PSF cores (as expected).  Hence, we transform the 2MASS J=6.20 magnitude of HD 181327 to the NICMOS F110W band using the CALCPHOT task in the STSDAS/SYNPHOT\footnote{http://www.stsci.edu/resources/software\_hardware/stsdas/synphot} synthetic photometry package.  Using an F6V input spectrum we find the  stellar flux density in NICMOS F110W band is 5.9 Jy, hence, the fraction of starlight scattered by the disk, {\it f}$_{scat}$, at {\it r}  $> $ 1\farcs2 is 0.170\% $\pm$ 0.015\% of the total stellar light.  We include in our error estimation the uncertainty in our flux density measure, and the absolute photometric calibration and transfer of the stellar flux density from 2MASS/J to NICMOS/F110W. This 1.1 $\mu$m scattering fraction is roughly comparable to the 0.25\%  thermal IR excess found from the $\geq$ 24 $\mu$m color-corrected {\it IRAS} and {\it Spitzer} flux densities ($\S 4$).

{\it  1.1 $\mu$m AXIAL RADIAL SURFACE BRIGHTNESS.} The SB was measured in radial ``strips'' along the disk major and minor axes, as well as in the direction along the image Y axis in the orientation of the ACS/HRC image (see Figure 6) with radial increments of 75.8 mas (one NICMOS camera 2  pixel) in both outward directions from the occulted star. At each radial increment the SB was determined from the median flux density measured in regions 1 pixel (along the radial) by 0$\farcs$5 (6.5 pixels, corresponding to a $\pm$ 10$\degr$ sector perpendicular to the radial at {\it R$_{peak}$}). The major axis SB profile is symmetrical about the star with a peak SB of $\sim$ 0.97 mJy arcsec$^{-2}$ at {\it r} = 1\farcs71.  In comparison, the asymmetrical SB peaks along the minor axis on opposite sides of the star ($\sim$ 1.37 and 0.79 mJy arcsec$^{-2}$) are strongly indicative of directionally preferential scattering along the line-of-sight.

{\it 1.1 $\mu$m DEPROJECTED AZIMUTHAL-MEDIAN RADIAL SB.} We measured the disk's 1.1 $\mu$m SB profile as a function of distance from the star with elliptical aperture photometry. In doing so we assumed intrinsic circular symmetry and a disk inclination of 31\fdg7 from a ``face-on'' geometry.  The measurement apertures were elliptical annuli of the same aspect ratio of the apparent disk ellipse (1:0.851) and co-aligned with the disk axes.  The elliptical annuli were 75.8 mas wide on the major axis, and were incremented in semimajor axis length by the same amount.   The deprojected radial SB profile medianed over all azimuth angles is shown in Figure 7.  The radial SB was found to peak at all azimuths at isophotal radii of 86.3 AU  from the star (i.e., 1$\farcs706$ if deprojected). The SB of the ring medianed over all azimuths at this peak radius was found to be 1.00 $\pm$ 0.07 mJy (F110W$_{mag}$ = 15.62 $\pm$ 0.07) arcsec$^{-2}$ (see figure 7), very closely corresponding to the ansal brightness of the ring on the major axis.  

{\it RING WIDTH.} On the azimuthal median, the measured FWHM of the deprojected ring (see figure 7) is 0\farcs538 ($\sim$ 29 AU). At all azimuth angles the SB of the ring declines from the 1$\farcs706$ (deprojected) radius of peak brightness more steeply toward the interior. The e$^{-1}$ SB falloff from the azimuthally medianed peak is 0\farcs334 (16.9 AU) inwardly and 0\farcs380 (19.2 AU) outwardly. In the region 1\farcs2  $< {\it r} <$ 5\farcs0 approximately  70\% of the  9.6 $\pm$ 0.8 mJy scattered light flux density at 1.1 $\mu$m is contained within annulus $\pm$ 18 AU in width centered on the brightest zone at  {\it r} = 86 AU from the star.

{\it 0.6 $\mu$m SURFACE BRIGHTNESS.} Because of the strong azimuthal and radial dependencies in the ACS PSF-subtraction
residuals, SB profiles in the region of the ring were measured from the ACS/HRC 0.6 $\mu$m image only at $r > 1\farcs4$  in azimuthal sectors within  $\pm$ 10$\degr$ of the image Y axis.  For reasons discussed in \S3.2, these are the only sectors in this annular zone where the flux density of the disk could be measured with reliable error estimates against the azimuthally anisotropic ``background'' of diffractive PSF residuals. To permit a direct comparison to the NICMOS 1.1 $\mu$m image, the ACS image was resampled to the NICMOS camera 2 pixel scale. The SB was measured in both directions from the occulted star along the ACS/HRC image Y axis in manner identical to the equivalent NICMOS measures.   The ACS/HRC image Y axis intersected the HD~181327 ring at 22$\degr$ from the disk major axis as determined from the NICMOS image (see figure 6). The 0.6 $\mu$m radial SB profile peaks at $\sim$ 1\farcs74 (at P.A. =  95$\degr$) and 1\farcs66 (at PA = 275$\degr$), with a mean SB of 0.57 $\pm$ $\lesssim $0.17 mJy (F606W$_{mag}$ = 16.9 $\pm$ $\lesssim$ 0.3) arcsec$^{-2}$, declining in intensity more rapidly inward than outward.

\subsection{Photometric Model of the 1.1$ \mu$m Ring-Like Disk}

Without consideration of the physical properties of the disk grains, we constructed a photometric model of the disk, predicated on the NICMOS 1.1 $\mu$m image, to determine the width of the ring, the grain scattering efficiency as a function of azimuth angle  (i.e., its scattering ``phase function'' in projection),  and to test the measurement of its isophotal radius. 

{\it FORM OF THE MODEL.}  We built a photometric model of the disk assuming an inclined, intrinsically circularly symmetric, ring\footnote{ We tested other functional forms, including model disks with uniform and radial power law surface brightness dependencies, but only the ring model provided a meaningful fit to the observational data. Both the uniform and radial power law SB models ``predicted'' excessive levels of scattered light interior to the peak radius of the ring that was recovered with equal fidelity after model implantation into the target-minus-PSF and PSF-minus-PSF images.  I.e., if the HD 181327 scattered light ring exhibited such a form, dust-scattered light interior to ring in excess of the declining levels  actually detected (eg., see fig. 7) would have been observable (and to even smaller radii due to more favorable image contrast) with statistical significance.}. Based on the results of the isophotal ellipse fitting (\S 5.1), we fixed the disk inclination to 31\fdg7, and orientation to 107$\degr$.  Allowing for the {\it r}$^{-2}$ dilution of the central starlight, the surface brightness of the ring was then parameterized by a set of four free parameters:  g$_{HG}$,  {\it R}$_{peak}$, SB($\theta$=0$\degr$,~{\it R}$_{peak}$), and {\it W},  as follows:

{\it AZIMUTHAL DEPENDENCE.}  We assumed a sky-plane projected azimuthal angle ($\theta$, measured in the plane of the disk) dependence  in the scattering efficiency (g$_{HG}$) of the disk grains with the line-of-sight to scattering-phase illumination angle, $\varphi$ = $ \cos^{-1}( \sin i \, \cos \theta)$. The  spatially resolved deprojected surface brightness, SB($\theta$,~{\it r}), with {\it r} measured in the plane of the inclined disk, is represented by a scattering phase function of the type suggested by  \cite{HG41}:

$$
SB(\theta,r) = SB(0\degr,R_{peak}) \times \left( \frac{1+{\mathrm g}_{HG}^2-2 {\mathrm g}_{HG} \sin i}{ 
1+{\mathrm g}_{HG}^2-2 {\mathrm g}_{HG} \cos \varphi}\right )^{1.5}
~~~~(1)~$$

For the inclination of the HD 181327 disk, scattering phase angles in the range of 58.3$\degr$ to 121.7$\degr$ are explored.

{\it RADIAL DEPENDENCE.} We modeled a radial SB dependence with  inward and outward exponential decays in SB from {\it R}$_{peak}$.  We characterize the ring width, {\it W}, as the distance between the radial points where the SB at any azimuth,  SB($\theta$,~{\it r}) declines to 
SB($\theta$,~{\it R$_{peak}$}) $\times$ e$^{-1}$, i.e.,  36.8\% of the peak SB 
 at the same azimuth. 

{\it CONVOLUTION.} The disk models  were convolved with the NICMOS camera 2 F110W point-source PSF for the SED of an F6 star using a noiseless, high fidelity, model PSF generated with the {\it HST} ``TinyTim''  optical model\footnote{http://www.stsci.edu/software/tinytim/tinytim.html} (version 6.3) as described by \cite{HOOK97}.

{\it FITTING PROCESS AND REGION.}  The set of parameters which best fit the observation for {\it i} = 31\fdg7, PA = 107$\degr$, was found by iterative $\chi^{2}$ minimization in observed minus model subtractions in radial zone 1\farcs2 $<$ {\it r} $<$ 3\farcs0.

The PSF subtraction residuals in the final image have complex spatial characteristics which increase in intensity with decreasing angular distance from the coronagraphically occulted star. Their amplitudes set the inner distance into which these observations can effectively probe. In the fitting process we exclude data at {\it r} $<$ 1\farcs2, as this region conveys little (or no) meaningful information.

At angular distances $>$ 1\farcs2, the instrumentally induced PSF-subtraction residuals are typically about a pixel in azimuthal extent but are correlated on much larger spatial scales radially.  These PSF ``tendrils'' impede the visibility of azimuthal variations in the ring surface brightness on comparable spatial scales.  To illustrate the underlying form of the ring, we removed these PSF subtraction residuals  by applying an elliptical low-pass filter to the image of the ring (figure 2C). We used a $\pm$ 9$\degr$ rotational ``boxcar'' kernel of 3.5 pixels width (i.e., 2.5 resolution elements at the radius suggested by isophote fitting), so only structures with approximately half that width are passed. Any ``real'' features of comparable azimuthal spatial frequencies are indistinguishable from PSF subtraction artifacts and also smoothed in this process. These high angular spatial frequency features are removed from the filtered disk image at the expense of a degradation in azimuthal spatial resolution.  The two bright spots in the NE sector of the ring (displayed as ``hard white'' in figure 2C) coincide with the location of the wider HST+NICMOS diffraction spikes in two of the four original (pre-combined) PSF subtracted images.  These  regions of enhanced brightness likely arise from larger spatial scale, but localized, imperfections in the PSF subtractions and are not intrinsic to the disk. For this reason, in the fitting process we also exclude these two regions in the NE quadrant of the ring  due to the likelihood of their arising from large spatial scale artifacts in the PSF-subtracted image.

\subsection{Model Fitting Results}

{\it PEAK RADIUS.} {\it R}$_{peak}$ was found to be 22.5 $\pm$ 0.4 pixels, in agreement with the radius found from fitting ring isophotes.  This corresponds to a physical radius of 86.3 $\pm$ 3.9 AU. 

{\it CHARACTERISTIC WIDTH.}  {\it W} was found to be 8.26 $\pm$ 0.9 pixels.  In this characterization, 
69.2\% of the light scattered by the dust originates in a 35.8 AU $\pm$ 4.0 AU wide annulus centered 86.3 AU $\pm$ 3.9 AU from the star (The 4.1\% uncertainty in the distance to HD 181327 from its parallax measure has been added in 
quadrature to the formal fitting error in expressing the uncertainty in {\it R}$_{peak}$ and  {\it W} in AU).

The  {\it W} = 35.8 AU $\pm$ 4.0 AU fit to the ring width (as characterized) is in  very good agreement with the estimate derived from the direct measures of the median radial profile of the deprojected disk.  Neither change significantly with azimuth when measured in 
9$\degr$ wide azimuthal sectors (in the absence of subtraction artifacts) around the ring.  
The large scale azimuthal uniformity of the FWHM$_{ring}$ suggests a constant product of the 
optical depth and grain albedo ($\tau$~$\times$~$\omega$) with azimuth, and 
perhaps, then, azimuthally undifferentiated grains.

{\it SCATTERING PHASE FUNCTION.}  The HD 181327 ring exhibits strong directionally preferential scattering with maxima and minima along the line of sight, i.e., in the direction of the ring minor axis, with g$_{HG}$ = 0.30 $\pm$ 0.03 as found in the fitting process.  (Discrete SB measurements along small azimuthal sectors at constant deprojected radii depart from this fit due to biasing by local PSF-subtraction residuals\footnote{The SB of the disk at  {\it R}$_{peak}$ was also discretely measured as a function of elliptical azimuth around the ring.  Measurements were made in ``face on projection'', with the disk surface brightness preserved in the geometrical transformation.  Circular photometric apertures of both 3.5 and 7 pixel diameters (one and two times the FWHM of the SB peak) were used to measure the SB every 9$\degr$ (40 measures), and every 18$\degr$ (20 measures) around the ring, respectively.  Aperture dilution of the non-uniform flux density distribution is less severe in discrete measurements with the smaller aperture, but these individual measures are more influenced by local subtraction residuals than with the larger aperture. The SB measures using both apertures were  in agreement with a g$_{HG}$ = 0.30 $\pm$0.03 scattering phase function.}). A photometric model of the disk, convolved with the NICMOS camera 2 PSF, g$_{HG}$ = 0.30, and {\it i} = 31\fdg7 from face-on (shown in Figure 2B) predicts ansal SBs of 0.863 mJy arcsec$^{-2}$ and minor axis SB peaks of 1.37 and 0.56 mJy arcsec$^{-2}$, each in agreement with the photometric measures (figure 6) within the 1~$\sigma$  measurement uncertainties assessed w.r.t. the azimuthal median.

\section{NICMOS + ACS}

The NICMOS and ACS instruments each better probed different spatial regimes of the HD~181327 circumstellar disk (for a comparative overview of the {\it HST} coronagraphs see \cite{SSH03}\footnote{http://nicmosis.as.arizona.edu:8000/POSTERS/AAS\_JAN2004\_CORON.jpg}). While the spatial domain of overlap with high photometric fidelity in the radial region of the debris ring is small, for the reasons noted in \S 3.2, together the NICMOS and ACS observations further inform on the global  structure and properties of the disk, and its environment.  The NICMOS imagery reveals the detailed structure of the disk and region interior to $\sim$ 3$\arcsec$, whereas the larger ACS field and deeper integrations reveal the faint outer nebulosity surrounding the ring at {\it r} $>$ 4\arcsec.  In the regions commonly sampled the nearly factor of 2 wavelength coverage afforded by the ACS/F606W  (0.6$\mu$m) and NICMOS/F110W  (1.1$\mu$m) filters provide rudimentary color information on the disk grains.

\subsection{[F606W] - [F110W] Color of the Disk at {\it R}$_{peak}$}
HD~181327 has a V-J (similar to [F606W]-[F110W] for an F5/F6V star) color index of $\sim$ +0.80. As noted in \S 5.2, the 0.6 $\mu$m SB of the disk could be measured reliably  only in small regions along the debris ring between 1\farcs4 and 2\farcs0 flanking the ACS/HRC image Y axis (indicated in figure 3A) at celestial position angles of 95$\degr$ and 275$\degr$, 22$\degr$ from the disk major axis (as shown in figure 6). Identical measures from the best-fit 1.1 $\mu$m scattered light model of the ring based upon the NICMOS image found SBs of 1.30 and 0.72 mJy arcsec$^{-2}$ at the same position angles, respectively.  The mean 1.1 $\mu$m SB in these regions corresponds to 15.61 $\pm$ 0.08 (F110W$_{mag}$ arcsec$^{-2}$).  Hence, in these regions where the disk-scattered light peaks radially, the disk grains appear significantly red with a [F606W] - [F110W] color index of +1.32 $\pm$ 0.3. Correcting for the color of the star, the disk grains in this region have an intrinsic [F606W]~-~[F110W] color index of +0.5 $\pm$ 0.3.

\subsection{The Outer Regions of the Disk}

Beyond {\it R}$_{peak}$, in the region of high photometric reliability of 100 - 200 AU ({\it r}~$\lesssim$~4$\arcsec$ in the deprojected image NICMOS image), the 1.1 $\mu$m SB declines as 15.75 x ${\it r}^{-5.2}$ mJy arcsec$^{-2}$ on the azimuthal median (see fig 8). Using the ring inclination and orientation derived from the NICMOS image, we similarly deproject the ACS image and find a very similar power law index with a monotonic decline in the 0.6 $\mu$m SB of 7.98 x {\it r}$^{-4.9}$ mJy arcsec$^{-2}$ in the same region (also shown in figure 8). The azimuthally medianed [F606W]~-~[F110W] color index of the disk throughout the 100 - 200 AU region is statistically identical to that found at the two measurable locations at 86 AU (\S 5.2).  Hence, from these limited data we see no evidence of radial differentiation in the color of the disk grains.

The ACS image reveals a diffuse low-SB nebulosity to a distance of $\lesssim$ 9$\arcsec$ from the star (figure 9, left panel), the inner portion  (beyond $\sim$ 4$\arcsec$) of which is only hinted at in the NICMOS image (figure 9, right panel).  The nebulosity seen in the ACS image, with a mean 0.6 $\mu$m SB in the 4$\arcsec$ $<$ {\it r} $<$ 9$\arcsec$ region of  $\approx$ 21.5 (F606W$_{mag}$) arcsec$^{-2}$, is azimuthally asymmetric with the brightest part in an $\sim$ 90$\degr$ sector to the North of the star and ``curving'' in the outer regions toward the east.  HST coronagraphic observations have revealed spiral structures in the circumstellar disks of the Herbig AeBe stars HD 100546 \citep{grady2001} and HD 141569A \citep{CLA03}. In the latter case, Clampin et al. postulated  that HD 141569's outer ``arms'' might have arisen from tidal interactions from a passing star, which could also be the causal mechanism for both the disk's stellar photocentric offset and its large azimuthal surface density variations.  Neither of these phenomena are seen in the HD~181327 disk. The southwest arm of the HD~141569A disk seems to be associated with its M-star companions, but no tidally interacting companions have been found for HD~181327 (see $\S$ 6.3). 
\citet{KAL02} have reported on the existence of Pleiades-like reflection nebulosities in the environments of Vega-like stars of spectral types K2V--B9.5V (i.e., similar to HD~181327), though an examination of {\it Spitzer} 24 and 70 $\mu$m images  of the field around HD~181327 ($\S$ 4) does not reveal any
emission structure the could be identified with the low-SB feature seen in the ACS 0.6 $\mu$m image.  Hence, the observations we report here, unfortunately, do not inform on the physical nature of the axially asymmetric nebular structure in proximity to HD~181327 that may, or may not, be physically associated with its disk. 

\subsection{Tests for Companionship}

In $\S 1$ we suggest that HD 181327 and the 10--30 Myr star HR 7329 \citep{Lowrance2000} are a common proper motion pair, strengthening the argument for HD181327's youth. Four point-like objects (stars) are seen in the more immediate vicinity of HD~181327 in both the NICMOS and ACS images (denoted {\it a}--{\it d} in Figure 9). 
Although stars {\it a} and {\it b} were within the 12\farcs6 $\times$ $12\farcs6$ SHARP-I \citep{NEU04} and 12\farcs7 $\times$ $12\farcs7$  SHARP-II \citep{CHA04}  survey fields, neither were detected in either study. We also searched for low-luminosity (sub-stellar) companion candidates in close angular proximity to HD 181327 using the NICMOS coronagraphic images. Following \citet {Schneider03} and \citet{Lowr05}, we implanted model ``PSF stars''(using noiseless TinyTim 6.3 PSFs \citep{HOOK97}) into a difference image made from of the two differentially-oriented coronagraphic images of HD~181327.  Even in the presence of the HD 181327 disk, at and interior to the radius of peak surface brightness, we found (similar to \citet{Wein99} in the case of HD 141569A) that  at this approximate age, brown dwarf or giant planet companions with masses $\gtrsim$ 10 M$_{jupiter}$ would be detectable in to an interior radius of $\approx$ 1$\arcsec$. To these approximate sensitivity limits, no close-proximity sub-stellar companion candidates were found.

The age of HD~181327, presumed $\sim$ 12 Myr as a member of the $\beta$ Pictoris moving group, could be further constrained by estimating the age of a low-mass (stellar) companion (e.g., by the presence of Li, H$_{\alpha}$ equivalent width, X-ray activity, etc.).  We tested the possibility of companionship of the four stellar-like objects appearing in the NICMOS and ACS images through differential proper motion (PM) measures over the 1.5 years between observations.  On the aggregate, the four had measured differential PMs of ($\Delta$RA, $\Delta$DEC) = (-42.3, +83.9) mas yr$^{-1}$ with a 1 $\sigma$ dispersion of $\pm$ (3.3, 16.0) mas yr$^{-1}$ w.r.t. HD~181327, suggestively indicative of the reflective motion of HD~181327 (the largest ``outlier'', star {\it c}, was discrepant from the ensemble mean by (4.8, 23.6) mas yr$^{-1}$). The proper motion of HD~181327 itself is given as (+24, -82) mas yr$^{-1}$ by \cite{peri97}.  Given the very good internal agreement in the measured PM of the four stars, the dispersion being $<$ (0.14, 0.20) $\sigma$ of the PM of HD~181327, all four -- unfortunately -- are very likely background stars\footnote{The magnitude of the apparent discrepancy in the reflective PM in RA of HD~181327 over the 1.5 years between the observations, assessed from the
 four star ensemble compared to its {\it a priori} measured value, is $\sim$ 28 mas (approximately 1/3 of a NICMOS camera 2 pixel). We attribute this discrepancy to an  (unresolved) inter-camera differential astrometric systematic uncertainty, likely due to differential camera distortion corrections and/or occulted target centration errors (as previously noted in \S3.2 for the ACS observation). We note that all four ``reference stars'' were spatially distributed within a single quadrant (to the west of HD~181327) due, in part, to the intersecting areas of the two camera fields.  A more isotropic distribution of reference stars (or a two-epoch observation with the same instrument) would have provided a better absolute astrometric solution. A fifth, brighter, star appears in the ACS field (to the east of HD~181327) only, so its differential PM could not be measured.}.

\section{DISK/GRAIN PROPERTIES}

\subsection{Inversion of the Scattered-Light SB Profile: Approach}

We modeled the scattered light surface brightness, {\it SB(x)}, of an axisymmetric, inclined, and optically thin disk, measured at a projected distance {\it x} from the star along the apparent major axis, following the methodology as  described by \cite{Aug06} such that:

$$
{SB(x) \over  \Phi^{*}} =
\int_{x}^{r_{max}}
\sigma_{sca}~\rho(r,z) \times 
{f(\varphi) + f(\pi-\varphi) \over r^{2} - x^{2}~\cos(i)} \times
{\sin(i)\over \sqrt{r^{2}-x^{2}}}
~{\mathrm r} dr
~~~ (2)
$$

where {\it i} is the disk inclination from pole-on, $\rho$({\it r, z}) is the disk
number density in the cylindrical frame centered onto the star
and attached to the disk, {\it r}$_{max}$ is the outer disk radius and $\Phi^{*}$
is stellar flux at Earth at the considered wavelength.
$\sigma_{sca}$ and {\it f}($\varphi$) are the scattering cross section of the grains
and the anisotropic scattering phase function at phase angle $\varphi$, respectively, 
both averaged over the grain size distribution.

In the case of HD~181327, the mean phase function {\it f}($\varphi$)  is observationally 
constrained and  can be approximated by a
Henyey-Greenstein type phase function with g$_{HG}$ =
0.3 $\pm$ 0.03 at 1.1 $\mu$m, as noted in \S 5.4. We also presume an 
axisymmetric disk predicated on the
observed bilateral symmetry of the scattered-light ring in the NICMOS image. 
Therefore, for a given vertical dust distribution, {\it Z}({\it r,z}), the product of the
midplane number density, $\rho$({\it r, 0})  =  $\rho$({\it r,z})/{\it Z}({\it r,z}),
and the mean scattering cross section, $\sigma_{sca}$, can be obtained
by directly inverting the observed surface brightness profile {\it SB}({\it x}).
We do so using the  technique detailed in \cite{Aug06} and discuss the properties of the 
dust ring as a function of {\it Z}({\it r,z}).

We characterize the vertical distribution with a three parameter function: 
{\it Z}({\it r,z}) = exp(-($|${\it z}$|$/{\it H}({\it r}))$^{\gamma}$); 
where {\it H} = {\it H}$_{0}$({\it r}/{\it r}$_{0}$)$^{\beta}$ 
is the vertical scale-height and {\it r}$_{0}$ is the observed radius
of the ring (86.3 AU).
We calculated the midplane number density for both exponential 
($\gamma$ = 1) and Gaussian ($\gamma$ = 2) profiles assuming
either linearly flaring ($\beta$ = 1) or a nearly flat ($\beta$ = 0.1) disks.
In each case, twenty scale-heights, 1 AU $\leq$ {\it H}$_{0}$ $\leq$ 20 AU were considered
corresponding to  0.012 $\lesssim  H_{0} /  r_{0} \lesssim$ 0.23 (see \S 7.3).

To compare the profiles, we introduce the disk surface density, $\Sigma$({\it r}),
that can be written
$$
\sigma_{sca}\Sigma(r) =
\int_{-\infty}^{+\infty}
\sigma_{sca}~\rho(r,z)~{\mathrm d}z = C_\gamma~H(r)~\sigma_{sca}~\rho(r,0) = \omega~\tau_{\bot}(r)~~~(3)
$$

where $\tau_{\bot}$({\it r}) is the vertical optical thickness at 1.1 $\mu$m and 
$\omega$ is the mean albedo in the NICMOS F110W passband averaged over  
the grain size distribution. {\it C}$_{\gamma}$ is a constant value depending upon $\gamma$
({\it C}$_{1}$  = 2, and {\it C}$_{2}$ = $\sqrt{\pi}$).

\subsection{Inversion of the Scattered-Light SB Profile: Results}

Figure 10 shows several $\sigma_{sca}$~$\Sigma$({\it r}) = $\omega$~$\tau_{\bot}$({\it r})
profiles obtained by directly inverting the averaged NICMOS F110W SB profile. For
sufficiently large values of {\it H}$_{0}$, the vertical optical thickness at  1.1 $\mu$m is independent of
the shape of  {\it Z}({\it r,z}) while the width of the ring\footnote{full width at e$^{-1}$ of the maximum}
of 43 $\pm$ 1 AU (FWHM  $\simeq$ 32.5 $\pm$ 0.5 AU) depends slightly on the vertical profile.
At large distances, the surface density profile falls off as $\sim$ {\it r}$^{-3}$, in agreement with the
observed  {\it r}$^{-5}$ SB profile. A surface density profile $\Sigma$({\it r}) $\propto$ {\it r}$^{-3}$
is fully consistent with dust grains supplied by parent bodies on circular orbits and placed in
orbits of high eccentricity by a drag force $\propto a^{2}/(r^{2}+z^{2})$ where
{\it a} is the grain size (e.g., see Eq. 3 and Fig 1 of \cite{Lec96}).
This model additionally requires a differential grain size distribution
that does not depart too much from a $a^{-3.5}$ power law, and is in good agreement
with the conclusions from the SED fitting (see \S 8).  Moreover, because the model
needs the dust dynamics to be largely unaffected by gas, the shape of the observed
dust profile may be indicative of low gas density as it has been similarly shown
for $\beta$ Pictoris \citep{theb05}.

As HD 181327 is a star of spectral type F5/F6V, radiation pressure contributes,
at least partly, to the total drag force on the grains. In that picture, the 
dust particles observed at distances larger than $\sim$ 100 AU most
likely originate from a belt of large solid bodies peaking
close to 87.6 AU.

The midplane optical thickness, $\tau_{\Vert}$, in the NICMOS F110W passband 
is related to the vertical thickness through

$$
 \omega~\tau_{\Vert}(r)  = \int_{0}^{r}~\sigma_{sca}~\rho(R,0)~{\mathrm
 d}R =
 \int_{0}^{r} {\omega~\tau_{\bot}(R) \over  C_\gamma~H(R)}~{\mathrm d}R
~~~~~~(4)
$$

 with $\tau_{\Vert}$ being the 1.1 $\mu$m optical thickness calculated from the region
between the inner edge of the disk to a distance {\it r}.  As shown in Figure 11, with
{\it H}$_{0}$ too small the disk appears marginally optically thick in the midplane.
Using the constraints on the grains for astronomical
silicates (\cite{Wein01}; \S 8), we estimate the mean albedo, $\omega$ to be $\approx$ 0.6.
Although the albedo is not well constrained, we can safely conclude
that {\it H}$_{0}$ must be $\gtrsim$ 4 AU ({\it H}$_{0}/${\it r}$_{0}$ $>$ 0.05)
to ensure the disk is optically thin in all directions at 1.1 $\mu$m.
When this condition is fulfilled, the ring vertical thickness,  $\tau_{\bot}$({\it r}), 
reaches a maximum of  $\sim$ 0.01 (assuming $\omega$ $\approx$ 0.6)
at 87.6 AU (Fig 10). Assuming a grain mass density of 3.5 g cm$^{-3}$ and a
{\it a}$^{-3.7}$ differential grain size distribution between 1 $\mu$m and 1 mm,
we infer a total dust mass of $\approx$ 4.1 M$_{Moon}$ ($\approx$ 0.05 M$_{\earth}$).

The collision time-scale may be estimated as suggested by \cite{BAC93},
$$
t_{coll} \simeq 
{\sigma_{sca} \over \pi <a^{2}> }~\times~
{r^{1.5} \over 2~\omega~\tau_{\bot}(r)~\sqrt{GM}}~~~~~(5)
$$

where 
$\pi<${\it a$^{2}$}$>$ is the mean geometric cross-section of the grains averaged
over the grain size distribution,
{\it G} is the gravitational constant, and {\it M} is the stellar mass ($\sim$ 1.35 {\it M}$_{\bigodot}$ 
for an F5/F6V star). The size distribution derived in \S~8 gives   
$\sigma_{sca}$/$\pi<${\it a$^{2}$}$>$ $\simeq$ 1.4. For such grains located within 
150 AU of the star, the collision timescale is $<$ 10$^{5}$ yr,  except very close to the inner
ring edge where the density is extremely low. At the position of the peak surface density,
{\it t$_{coll}$} $\approx$ 10$^{4}$ yr (hence about
three orders of magnitude smaller than the star's age). The ring
is then collision-dominated and the smallest of the released
grains can populate the outer disk thanks to radiation pressure 
commensurate with the {\it r}$^{-3}$ surface density profile at large distances.
The short collision time-scale implies that any dynamical process
for for the dust grains lasting longer than 10$^{4-5}$ yr can thus be
neglected. This is the case of Poynting-Robertson drag for which the depletion
timescale at 87.6 AU for just-bound grains compares to the stellar age
(i.e., $\sim$ 10 Myr). I.e., dust cannot populate the disk interior to the
ring due to Poynting-Robertson drag.

\subsection{Scattered-Light Models and Fits}

Synthetic scattered-light images of the HD 181327 circumstellar disk were generated employing the three dimensional radiative transfer code, MCFOST, described by \cite{pinte06}.  The product of density, $\rho(r, z)$, and scattering cross section, $\sigma_{sca}$, resulting from the inversion of the surface
brightness profiles (\S\S 7.1 \& 7.2) were used assuming a Henyey-Greenstein scattering
phase function with g$_{HG} = 0.3$ (\S 5.4).  Photon packages escaping the system with inclination angles between 30\fdg5 and 32\fdg5 were recorded in the synthetic images. Representative model disks for a subset of characterizing 
parameters $\alpha, \beta,$ and $H_{0}$, corresponding to several of the profiles discussed
in \S7.2 (and presented in figures 10 and 11), are shown in figure 12.

The spatially resolved flux densities predicted by the MCFOST scattered-light models were fit to the  NICMOS image in the annulus of high SNR ($r \approx 1\farcs3 - 2\farcs3$) enclosing the observed scattered-light ring. Each of the previously discussed models, tiling the parameter spaces illustrated in Figure 12, were fit to the NICMOS image by $\chi^{2}$ minimization, scaling the intensities of the predicted scattered-light patterns to the observed spatially-resolved flux densities.  A representative example (the $\beta = 1, \gamma = 2, H_{0}$ = 1AU case), and its fit to the NICMOS image after $\chi^{2}$ minimization, is shown in Figure 13.  Within each ($\beta, \gamma$) ``family''  of models, the reduced $\chi^{2}$ estimation was relatively flat for $H_{0} < 8$ AU. In \S 7.2 we separately concluded that  {\it H}$_{0}$ must be $\gtrsim$ 4 AU ({\it H}$_{0}/${\it r}$_{0}$ $>$ 0.05), hence we suggest 4 AU $\lesssim H_{0} \lesssim 8$ AU.

\section{GRAIN SIZE DISTRIBUTION}

Both the ACS (0.6 $\mu$m) -- NICMOS (1.1 $\mu$m) color of the HD~181327 disk and
the measured scattering asymmetry factor (g$_{HG}$ = 0.3 at 1.1 $\mu$m) indicate that
the  scattered light images are dominated by sub-micron grains. 
The grain properties that may explain the scattered light observations
should also be consistent with other
independent constraints for the grains. In particular, grains
at the observed radius of 86.3 AU should produce an infrared signature 
consistent with the measured long-wavelength excesses and
the grains properties should be consistent with basic dynamical considerations. Because
HD~181327 is an F5--F6 type star, the radiation pressure force on the
grains is expected to be significant and can expel sufficiently
small grains from the disk on very short time-scales.

We have assumed a unity normalized differential grain size distribution 
($\int_{a_{min}}^{a_{max} }$d{\it n}({\it a}) = 1) 
of the form d{\it n}({\it a}) $\propto$ {\it a}$^{\kappa}$~d{\it a}
between a minimum grain size, {\it a}$_{min}$, and
a maximum grain size,  {\it a}$_{max}$.  We thus search for 
({\it a}$_{min}$, $\kappa$) pairs consistent
with the four constraints of color, g$_{HG}$, IR excess, and blow-out
size limit  (thus, corresponding to the depletion timescale), identified above.
In the following, the maximum grain size will be assumed to
be large enough to not affect the results. 
The grain optical properties are calculated with the Mie
theory valid for hard spheres and for astronomical silicates (optical
constants from \cite{Wein01}).

The measured mean scattering asymmetry factor, g$_{HG}$  = 0.3 $\pm$ 0.03 in the F110W passband,
then reads

$$
{\mathrm g}_{HG} (a_{min}, \kappa) = 
\int_{a_{min}}^{a_{max}}
{\mathrm g}(a)~\pi~a^{2}~Q_{sca}~{\mathrm d}n(a)~/~\sigma_{sca}~~~~~(6)
$$

where {\it Q}$_{sca}$ is the dimensionless scattering efficiency,
$\sigma_{sca}$ = $\int_{a_{min}}^{a_{max}}$
$\pi${\it a}$^{2}~${\it Q}$_{sca}$~d{\it n}({\it a})
and g({\it a}) is the grain-size dependent scattering asymmetry factor
at 1.1 $\mu$m. As the disk is optically thin, the measured disk color is
dependent upon the ratio of scattering cross sections and phase functions
at the two observed wavelengths.  The mean scattering asymmetry factor
at 0.6 $\mu$m, unfortunately, could not be measured from the ACS image.
We, therefore, have made the rough assumption that the phase function in
the ACS passband is comparable to that found from the NICMOS (1.1 $\mu$m)
observations. In that case, the measured color directly corresponds to the
mean scattering cross section ratio: [F606W]$-$[F110W] =
2.5 log $_{10}$($\sigma_{sca}$[F606W]/$\sigma_{sca}$[F110W])
= +0.5 $\pm$ 0.3.

To further constrain the grain size we used the surface
density profile obtained in \S 7 to calculate infrared flux
densities in the $\lambda > 12~\mu$m IRAS\footnote{The 12 $\mu$m emission detected
by IRAS arises from the stellar photosphere, not the disk.} and Spitzer/MIPS bands (Table 2) as a functions of
{\it a}$_{min}$ and $\kappa$ using the \cite{Aug99} model. Fitting the SED leads to reduced $\chi^{2}$ 
contours in the ({\it a}$_{min}$, $\kappa$)
parameter space as illustrated in Figure 14. The best fit is
obtained for minimum grain sizes close to 1 $\mu$m, in very good agreement with
the radiation-pressure induced minimum grain size. The
SED fitting also reduces the range of possible $\kappa$ values, 
the best fit being obtained for a power law index of $\approx$ -3.7,
quite close to the theoretical -3.5 power law index that holds for systems in collisional equilibrium
\citep{Doh69} (dashed line in Figure 14).

The ACS-NICMOS disk color is reproduced with only slightly smaller (0.4 $\lesssim$ {\it a}$_{min}$  $\lesssim$ 0.6 $\mu$m) grains
than those with sizes that best  fit the thermal SED 
(1 $\lesssim$ {\it a}$_{min}$  $\lesssim$ 2 $\mu$m). Nevertheless, these grains remain  too large to reproduce
the measured asymmetry factor g$_{HG}$ (requiring   {\it a}$_{min}$ $\lesssim$  0.1 $\mu$m grains in our
formulation). In the frame of the
Mie theory, the observed mean asymmetry factor can indeed
only be reproduced when scattering is dominated by very small
grains which is achieved when both {\it a}$_{min}$ and $\kappa$ 
are extremely small.  This inconsistency may relate to the fact that while we have calculated
the asymmetry factor for hard silicate spheres (a simplifying assumption inappropriate for
grains with more complex structures), the scattering phase function is
very sensitive to the shape and structure of the grains.  Additionally, the HD~181327 disk may
possess disk grains that: (a) originate in more than just a  single population, (b) are radially
segregated in distance from the star, and (c) have a grain-size distribution not well-represented by a 
single power-law.

The grains which are predominantly responsible for HD~181327's thermal infrared excess may not be those which give rise to the circumstellar light scattered at 0.6 to 1.1 $\mu$m.  In the case of the debris system circumscribing HR~4796A \citep{schn4796}, \cite {Aug99} suggested a two-component dust population with 
large grains of ISM-like composition (moderately porosity amorphous silicates) at the radius of the scattered-light disk, and high porosity crystalline grains close to the star. The thermal emission from HD~141569's circumstellar disk was shown by \cite{Marsh} to arise from within 1$\arcsec$ (100 AU) of the star  from 12 -- 21 $\mu$m imaging, while light scattered by the disk at 1.1 and 1.6 $\mu$m (\cite{Wein99}; \cite{Aug99b}) originates (and extends) significantly further out.  While HD~181327's thermal SED is not inconsistent with the possibility of ISM like grains (e.g., $\kappa$ $\approx$ -3.5 with {\it a}$_{min}$ $<$ 1  $\mu$m ; \citet{mathis}) in the scattered-light debris ring, the presumption of a single population of hard, spherical, silicate grains would be inapplicable for grains which have grown (evolved) from ISM-like structures, and may be spatially (radially) segregated in distance from the star.

\section{DISCUSSION}

The inferred spatial distribution of dust detected about stars, from thermal infrared excesses (as measured from {\it IRAS} and {\it Spitzer}), often predicts central clearings or ``holes'' to explain the observed spectral energy distributions.  Scattered-light imagery of circumstellar debris directly provides  constraints on the spatial distribution of dust independent of the dust grain temperature in regions sufficiently far from the central star which can be instrumentally probed with high spatial resolution.  The central regions of dusty circumstellar disks that are nearly edge on to the line-of-sight (e.g., $\beta$ Pictoris, AU Microscopii and HD~32297) remain hidden at optical and near-IR wavelengths due to the opacity of the dust.  The inner regions of debris disk systems which are more inclined to the line-of-sight, such as HD~181327, are revealed directly and give  support to the interpretation of inner holes from analyses of debris disk SEDs. While only a very small number of light-scattered images of circumstellar debris disks exist, the growing number allows inter-comparison of the disk characteristics with stellar type and age (when sufficiently constrained; see Table 3 and Figure 15).  With the notable exception of $\alpha$ PsA, which thus far has only been observed in scattered light with {\it HST}/ACS, all of the debris disks listed in Table 3 have been observed with scattered-light imagery in the optical and the near-infrared. Very recently, two newly discovered, scattered-light disks around older (0.3 -- 1 Gyr) stars, HD~53143 and HD~139664 \citep{KAL06}, have been reported, adding importantly to this still very limited sample.  

Of the debris disks imaged to date, the HR~4796A ring-like disk is, in many respects, morphologically most similar to HD~181327's. Both scattered-light disks appear as relatively narrow annuli, have similar near-IR scattering fractions and thermal IR excesses, and physically are of similar size (70 and 86 AU in radii, respectively).  Both stars, while not coeval (HR~4796A being a member of the TW Hya association, and HD~181327 in the $\beta$ Pictoris moving group)  are of similar ages ($\sim$ 10 Myr), and spectral types.  

Directionally preferential scattering, symmetrical about the disk minor axis (i.e., with respect to the line-of-sight), is seen in the HD~181327 disk. Such non-isotropic scattering might be expected from circumstellar particles evolved from a primordial ISM-like population (comparable in size to, or larger than, the  1.1 $\mu$m wavelength of the observation), and similar behavior is seen in the HR~4796A debris ring, and in $\alpha$ PsA's and HD~107146's as well.  

These four ring-like debris disks vary by factors of $\sim$ 2 in radius (from the already noted smaller HR~4796A and HD~181327 to the larger HD~107146 (133 AU; \cite{ardila04}) and $\alpha$ PsA (145 AU; \cite{KAL05}), and $\sim$ 4 in ring width-to-radius (w:r) ratio (from $\sim$ 0.17, for both HR~4796A \citep{schgil} and  $\alpha$ PsA, to 0.64 for the very broad HD~107146 ring, and HD~181327 in between at 0.38). The disks about HR~4796A and  $\alpha$ PsA, both early A stars, are most disparate in radius but most similar in w:r ratios, suggesting a lack of simple scalability in ring geometries conflated (if not determined) instead by  dynamical and kinematic processes in play. For example, in the case of HR~4796A, the outer radius of its disk may be truncated by the dynamical influence of its M-star companion (e.g., \cite{arty94}). The sharp inner and more shallow outer edges of the illumination-corrected scattered-light radial profiles of HD~181327 (Fig. 7) and  $\alpha$ PsA (figure 3 of \cite{KAL05}) are remarkably similar, suggestive of similar clearing and/or confinement mechanisms for the disk grains.

The photocenter of the $\alpha$ PsA's dust ring is asymmetrically offset from the position of the central star. As suggested by \cite{KAL05}, such an asymmetry may implicate the existence of one (or more) planetary perturbers. The HR~4796A debris ring exhibits an $\sim$~20\% SB asymmetry at the ring ansae at near-IR and optical wavelengths \citep{sch04} (also seen in it's mid-IR thermal emission as reported by
\cite{telesco}), with no photocentric offset in the position of the star relative to the ring ansae detectable in the scattered-light observations. Such a non-axisymmetric SB asymmetry about the ring minor axis could also arise from a local enhancement (or depletion) in the dust density distribution due to perturbations from unseen, co-orbital, planetary-mass bodies (e.g., \cite {ozer}).  Though, alternative mechanisms, dependent upon the amount of residual gas in the system, could account for the ring structure itself (e.g, \cite{klar}; \cite{tag}) without invoking planetary-mass companions.  

The HD~181327 ring morphology is bi-axially symmetric, centered on the star, and devoid of any readily apparent azimuthal  SB anisotropies aside from those dependent upon scattering phase angle.  Such a ring structure could arise from gas drag in the absence of planets, though the gas-to-dust ratio in the HD~181327 disk is, currently, unknown.

\section{SUMMARY}

We have imaged a circumstellar debris disk about the $\sim$ 12 Myr old F5/F6V star
and $\beta$ Pictoris moving group member HD 181327 in 1.1~$\mu$m and 0.6 $\mu$m light scattered by the disk grains.  
Our 1.1 $\mu$m PSF-subtracted  NICMOS coronagraphic observations reveal a bilaterally symmetric ring-like structure, 
31\fdg7 $\pm$ 1\fdg6 from face-on with a  major axis PA of 107$\degr$ $\pm$ 2$\degr$ east of north.
We estimate the 1.1~$\mu$m disk flux density in the region 1\farcs2 $<$ {\it r} $<$ 5\farcs 0 to be 
9.6 $\pm$ 0.8 mJy (0.17 \% of the total stellar light) with $\sim$ 70\% of the disk flux density contained within a  36 AU wide annulus
centered on the brightest zone at  {\it r} = 86 AU from the star, well beyond the blackbody characterized 22 AU radius of the thermally emissive dust detected in the {\it IRAS} and {\it Spitzer}/MIPS bands.  The scattered-light disk appears to be centrally
cleared (to within 1\farcs2 of the star, which is the limiting region which can be effectively probed
with our NICMOS observations) and centered on the star.  

The surface brightness of the ring of debris-scattered light is well represented by a \cite{HG41} type scattering phase function with g$_{HG}$ = 0.3.  Modeling the disk, we find
a vertical scale height $H_{0}$ = 5--10\% of  the 87.6 AU radius of the maximum surface density at that radius.
The observed dust ring explains the  thermal IR excess emission above the stellar photospheric level measured by {\it IRAS} and {\it Spitzer} provided that $a_{min} \gtrsim 1~\mu$m with a grain size distribution $\propto a^{-3.7}$, presuming hard, spherical, silicate grains.  With these grain parameters, and surface density profiles derived by direct inversion of the SB profiles obtained from the NICMOS imagery, we infer a total dust mass of $\approx$ 4 M$_{Moon}$ ($\approx$ 0.05 M$_{\earth}$) in the collision-dominated ring. Such grains, however, fail to reproduce the  g$_{HG}$ = 0.3 scattered-light asymmetry factor, suggestive of a more complex (possibly mixed) population of disk grains.  To better constrain the possible grain sizes and their radial distributions, in {\it HST} cycle 15  we  will obtain 2 $\mu$m coronagraphic imaging polarimetry  observations with NICMOS (a new, recently enabled, observing mode; \cite{Hines04}) of the HD 181327 disk (GO program 10847).

Additional circumstellar  light is seen from 2\arcsec--4\arcsec~(100--200AU) in both 1.1 $\mu$m NICMOS and 0.6 $\mu$m ACS imagery
declining in SB as $\sim$ {\it r}$^{-5}$ in both bands. 
The SB profile beyond 100 AU is consistent with the blow-out time scale for grains in the inner disk by 
radiation pressure, collisional grains with a size distribution as discussed, and low gas pressure in the disk.
A low SB diffuse outer halo (V $\approx$ 21.5 mag arcsec$^{-2}$) is seen
in the ACS image extending as far as $\sim$ 9\arcsec~(450 AU) to the north of the star.

\acknowledgments
We are grateful to F.~J.~Low and P.~S.~Smith of the Spitzer GTO (PID 72) team for providing  MIPS imaging photometry of HD~181327. We thank Elizabeth Stobie (NICMOS Project, University of Arizona) for her significant contributions
to this program through her development, implementation, enhancement and support of analysis
software used throughout the course of this study. We appreciate the commentary received from Eric Mamajek, particularly in regard to the age estimation for HD 181327 beyond its ``guilt by association'' as a presumed member of the $\beta$ Pictoris moving group, as reflected in our discussion in $\S\S 1$ and 6.3 of this paper. This investigation was based on observations made with the NASA/ESA Hubble Space Telescope,
obtained at the Space Telescope Science Institute (STScI), which is operated by the Association of Universities
for Research in Astronomy, Inc., under NASA contract NAS 5-26555.  These observations are associated
with programs GO/10177 and GTO/9987. Support for program GO/10177 was provided by NASA through
a grant from STScI. S. Wolf was supported by the German Research Foundation (DFG)
 through the Emmy Noether grant WO 857/2-1. J.C.Augereau, C. Pinte and F. M\'enard acknowledge financial support from the {\sl Programme National de 
Physique Stellaire} (PNPS) of CNRS/INSU, France. We acknowledge the use of the SIMBAD database.




\clearpage

\begin{deluxetable}{llccllccl}
\tabletypesize{\scriptsize}
\tablecaption{NICMOS OBSERVATIONS OF HD 181739 and PSF Refrence Stars}
\tablewidth{0pt}
\tablehead{
\colhead{Visit\tablenotemark{a}} & \colhead{Star\tablenotemark{b}} & \colhead{Orient\tablenotemark{c}} & \colhead{Type\tablenotemark{d}} & 
\colhead{Filter\tablenotemark{d}} & \colhead{Readout Mode\tablenotemark{d}} & 
\colhead{ExpTime\tablenotemark{e}} & \colhead{Exp\tablenotemark{f}} &
\colhead{Datasets\tablenotemark{g}}
}
\startdata
67 & HD 181327 & 37.570 & ACQ  & F180M & ACQ (ACCUM) & 0.243s  &  2 & N8ZU67010 \\
   &     &   &CORON & F110W & STEP32/NSAMP14 & 224s & 3 & \\
   &     &  &CORON & F110W & STEP32/NSAMP10 & 96s & 1 & \\
   &     &   &DIRECT& F110W & SCAMRR/NSAMP11 & 2.03s& 2 & \\
68 & HD 181327 & 66.970 & ACQ  & F180M & ACQ (ACCUM) & 0.243s  &  2 & N8ZU68010 \\
   &     &   &CORON & F110W & STEP32/NSAMP14 & 224s & 3 & \\
   &     &  &CORON & F110W & STEP32/NSAMP10 & 96s & 1 & \\
   &     &   &DIRECT& F110W & SCAMRR/NSAMP11 & 2.23s& 2 & \\
83 & HD 4748 & 278.970 & ACQ  & F171M & ACQ (ACCUM) & 0.309s  &  2 & N8ZU83010 \\
   &     &   &CORON & F110W & STEP32/NSAMP12 & 160s & 1 & \\
   &     &   &CORON & F110W & STEP32/NSAMP14 & 224s & 10 & \\
   &     &   &DIRECT& F110W & SCAMRR/NSAMP10 & 2.03s& 8 & \\
84 & HIP 7345 & 198.398 & ACQ  & F187N & ACQ (ACCUM) & 0.763s  &  2 & N8ZU84010 \\
   &     &   &CORON & F110W & STEP32/NSAMP10 & 96s & 1 & \\
   &     &   &CORON & F110W & STEP32/NSAMP14 & 224s & 10 & \\
   &     &   &DIRECT& F110W & SCAMRR/NSAMP10 & 2.03s& 8 & \\
\enddata
\tablenotetext{a}{{\it HST}$/$GO 10177 Visit ID number}
\tablenotetext{b}{ 2MASS mags from \cite{2MASS}, Spectral types from \cite{peri97}: \\
 ~~~~~[HD 181327; VT = 7.095 $\pm$ 0.007, J = 6.20 $\pm$ 0.02, H = 5.98 $\pm$ 0.04,  K = 5.91 $\pm$ 0.04, F5/F6V],\\
 ~~~~~[HR 4748; J = 5.52 $\pm$ 0.02, H = 5.58 $\pm$ 0.03,  K = 5.54 $\pm$ 0.02, B8V],\\
 ~~~~~[HIP 7345; J = 5.49 $\pm$ 0.02, H = 5.53 $\pm$ 0.02,  K = 5.46 $\pm$ 0.02, A1V].
}
\tablenotetext{c}{Position angle (degrees) of image +Y axis (East of North).}
\tablenotetext{d}{See Noll et al. (2004) for NICMOS image/readout modes and filters.}
\tablenotetext{e}{Exposure time for each exposure (before combining).}
\tablenotetext{f}{Number of exposures}
\tablenotetext{g}{{\it HST} archive data set ID}
\end{deluxetable}

\begin{deluxetable}{ccccc}
\tabletypesize{\scriptsize}
\tablecaption{Spitzer/MIPS  \& IRAS Thermal IR Flux Denisties}
\tablewidth{0pt}
\tablehead{
\colhead{Wavelength ($\mu$m)\tablenotemark{a}} & \colhead{Source} & \colhead{Flux Density (Jy)\tablenotemark{b}}& Uncertainty &\colhead{Color-Corrected\tablenotemark{c} (Jy)}
}
\startdata
12&IRAS& 0.164&10\%&0.114 $\pm$ 0.011\\
23.68&MIPS&0.220&4\%&0.223 $\pm$ 0.009\\
25&IRAS&0.248&8\%&0.306 $\pm$ 0.024\\
60&IRAS&1.86&4\%&1.96 $\pm$ 0.08\\
71.42&MIPS&1.58&7\%&1.73 $\pm$ 0.12\\
100&IRAS&1.72&12\%&1.70 $\pm$ 0.20\\
155.9&MIPS&0.750&12\%&0.766 $\pm$ 0.090\\
\enddata
\tablenotetext{a}{Weighted mean wavelengths for MIPS; see http://ssc.spitzer.caltech.edu/mips/dh/}
\tablenotetext{b}{IRAS  flux densities from the FSC}
\tablenotetext{c}{for 70K blackbody grains in all bands except {\it IRAS} 12 $\mu$m: for 6450K photosphere. See:\\
~~~~~http://irsa.ipac.caltech.edu/IRASdocs/exp.sup/ch6/tabsupC6.htm\\
~~~~~http://ssc.spitzer.caltech.edu/mips/dh/mipsdatahandbook3.2.pdf}

\end{deluxetable}
\clearpage

\begin{deluxetable}{cccccccc}
\tabletypesize{\scriptsize}
\tablecaption{Circumstellar Debris Systems Imaged in Scattered Light}
\tablewidth{0pt}

\tablehead{
\colhead{Star} & 
\colhead{Spec} &
\colhead{Age (Myr)} &
\colhead{r(AU)\tablenotemark{a}} &
\colhead{{\it f}$_{\rm nir}$(r$>$0\farcs3)} &
\colhead{L$_{\rm ir}$/L$_{*}$} &
\colhead{Assymetry} &
\colhead{Image}
}

\startdata
AU Mic$^{b}$&M0&12--20&210&...&0.0004&Wing Tilt&Fig 15a\\
$\beta$ Pic$^{c}$&A5&12--20&$~$1600&0.003&0.0015&many&Fig 15b\\
HD 32297$^{d}$&A0&$\le$10 ?& $>$400&0.0033&0.0027&SB/W. Tilt&Fig 15c\\
HR 4796A$^{e}$&A0&8&70&0.0024&0.005&Azim.&Fig 15d\\
HD 181327&F5/6&12&86&0.0017&0.0025&No, g$\approx$0.3&Fig 15e\\
HD 141569A$^{f}$&HAeBe/B9.5&5&400&0.0025&0.0084&Yes&Fig 15f\\
HD 107146$^{g}$&G2&30-250&170&1.4x10$^{-4}$&0.001&No, g$\approx$0.2&Fig 15g\\
$\alpha$ PsA$^{h}$&A3&200&140&10$^{-6}$&5x10$^{-5}$&Offset, g$\approx$0.3&Fig 15h\\
\enddata
\tablenotetext{a}{r$_{\rm AU}$ reported is scattered-light maximum extent detected for near edge-on disks (and HD 141569A), radius of peak surface brightness for ring-like disks (HR 4796A, HD 181327, and  $\alpha$ PsA).}
\tablenotetext{b}{AU Mic: \cite{KAL04}.}
\tablenotetext{c}{$\beta$ Pic: \cite{kal95}, {\it f}$_{\rm nir}$ estimated from \cite{KAL00}, L$_{\rm ir}$/L$_{*}$  from \cite{DEC03}.}
\tablenotetext{d}{HD~32297: \cite{SCH05}.}
\tablenotetext{e}{HR~4796A: \cite{schn4796}.}
\tablenotetext{f}{HD 141569A: \cite{Wein99}, anisotropies see \cite{mou01} \& \cite{CLA03}.}
\tablenotetext{g}{HD~107146: \cite{ardila04}, g at 0.8 $\mu$m.}
\tablenotetext{h}{$\alpha$ PsA: \cite{KAL05}, L$_{\rm ir}$/L$_{*}$  from \cite{DEC03}.}
\end{deluxetable}
\clearpage


\clearpage
\begin{figure}
\epsscale{1.0}
\plotone{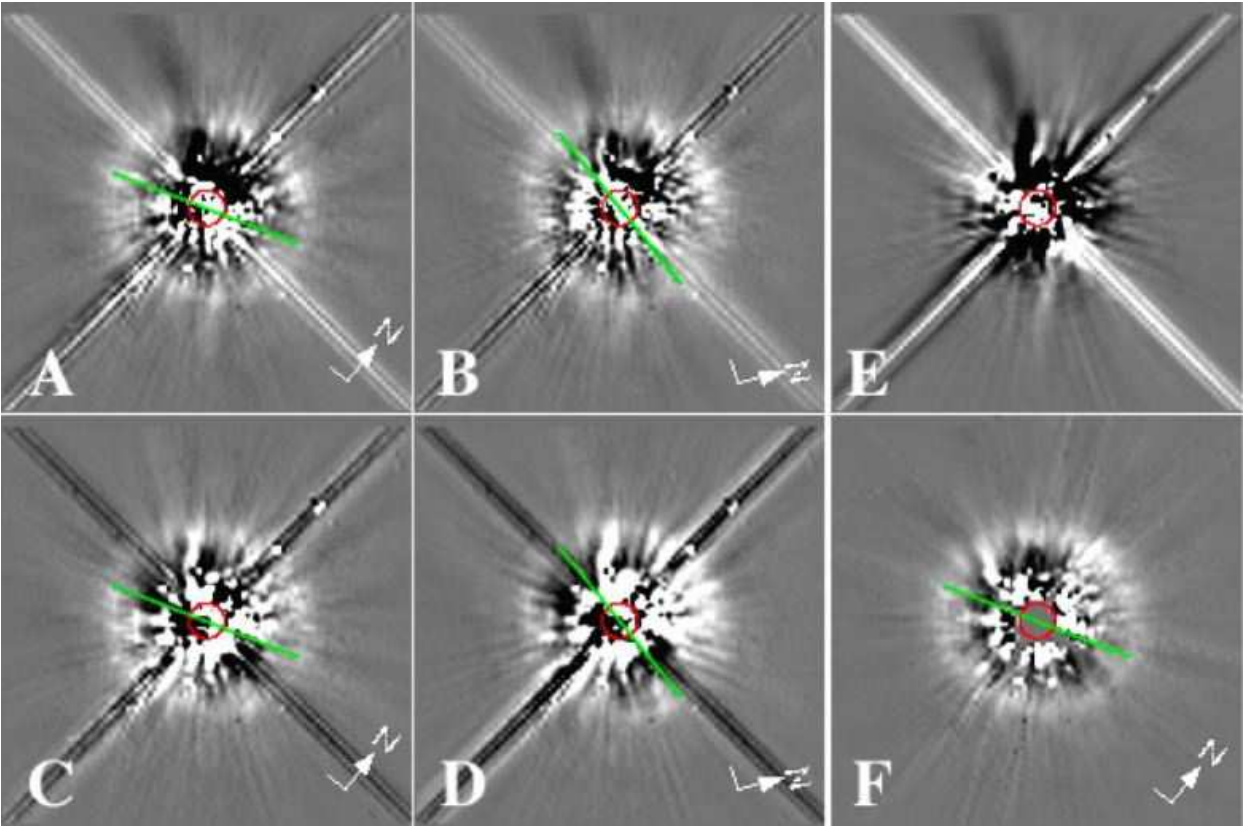}  
\caption{Fig. 1 --  
The robustness of NICMOS PSF-subtracted coronagraphy in revealing the HD 181327 circumstellar debris ring is illustrated by these individual (A-D) representative 
astrometrically registered, and flux-scaled difference images, using two different PSF template stars (HIP 7345 in panels A \& B; HIP 4748 in panels C \& D) with HD 181327 observed at two field orientations differing by 29\fdg9 (absolute orientations 
w.r.t.~North as indicated).  Panel E shows a same-scaled and registered  PSF-minus-PSF subtraction of the two template stars. All images are shown with a symmetrical bi-polar stretch of $\pm$ 10 counts sec$^{-1}$ pixel$^{-1}$ (F110W surface brightness of $\pm$ 2.19 mJy arcsec$^{-2}$) to show the nature (and intensity) of the PSF subtraction residuals.  Differential diffractive and scattering artifacts remain spatially fixed on the detector with the spacecraft re-orientation, but vary in intensity and phase due, primarily, to spacecraft ``breathing''  \citep{schn01}.  Note for example the diagonal diffraction spikes from the HST secondary mirror support, and a commonly occurring diffuse coronagraphic scattering feature (known as ``the smokestack'', slightly curving to the upper 
left, dark in panels A, B \& E) which appears  in conjunction with a more intense ``finger'' closer to and above the coronagraphic hole (red circles), particularly bright in panels C \& D.  Panel F combines images A-D, medianed with equal weight after masking the diffraction spike artifacts in the individual images and rotating all to the same celestial orientation.  All panels are $7\arcsec \times 7 \arcsec$. Green lines indicate the disk major axis.
}
\end{figure}

\clearpage
\begin{figure}
\epsscale{1.00}
\plotone{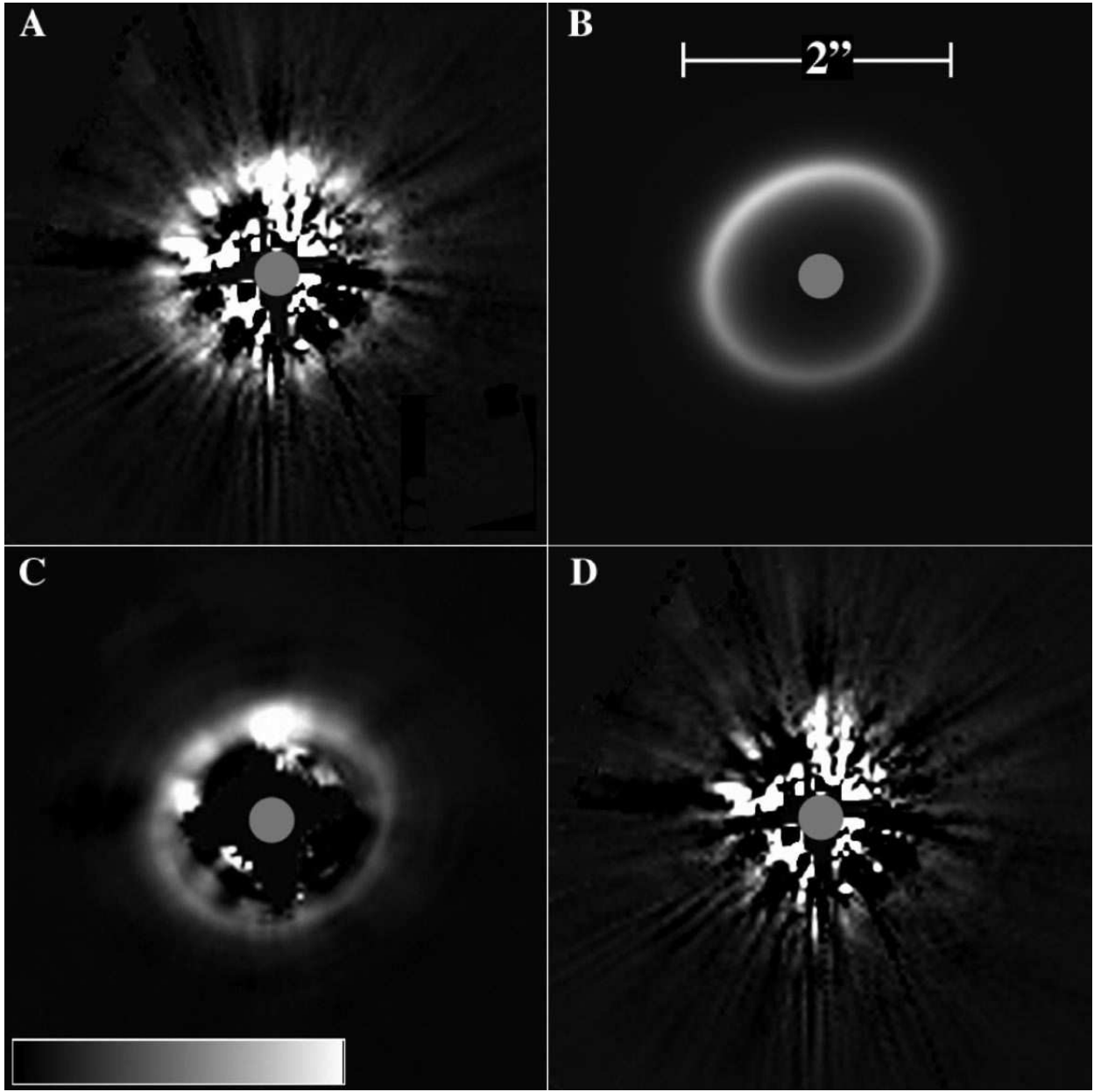}  
\caption{A) NICMOS 1.1 $\mu$m two-orientation combined PSF-subtracted coronagraphic image of the HD~181327 debris ring (same image data as Figure 1F). B) Best fit photometric model, as discussed in \S 5.4 of the text. C) Observed image (from panel a) after low-pass filtering with a 9$\degr$ elliptical ``boxcar'' kernel.  D)  Residuals after subtracting (B) from (A).  All panels shown at the same linear display stretch: -0.5 (black) to +10 (white)  counts sec$^{-1}$ pixel$^{-1}$ (-0.11 to +2.19 to  mJy arcsec$^{-2}$) and scale (as indicated).  Small ($r = 0\farcs3$) gray circles are centered on the location of the star.}
\end{figure}

\clearpage
\begin{figure}
\epsscale{1.00}
\plotone{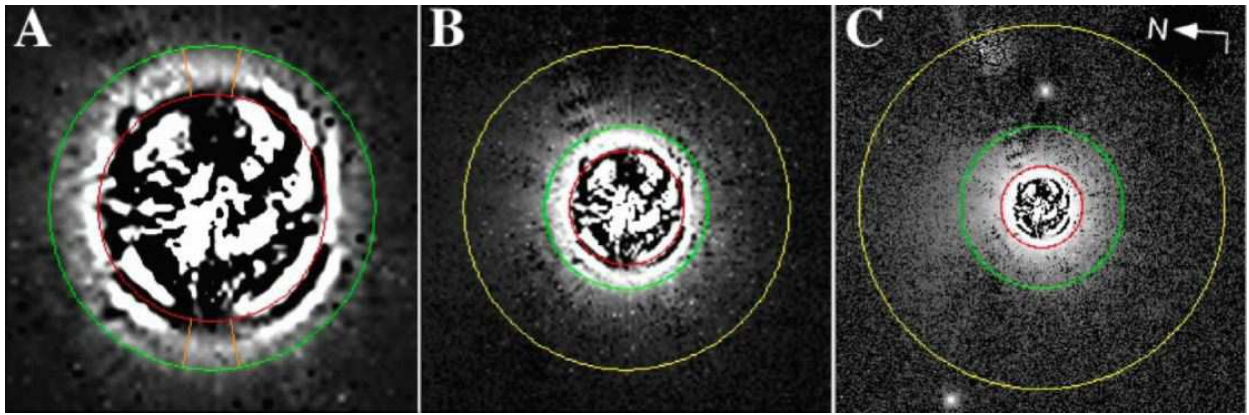}  
\caption{ACS 0.6 $\mu$m PSF-subtracted coronagraphic image of  HD~181327.~~A) $2\farcs5$ x $2\farcs5$ field, $r = 1\farcs4$ (red), $r = 2\farcs0$ (green), -0.1 to +1.5 e$^{-}$~s$^{-1}$~pixel$^{-1}$ linear display. The disk is reliably seen only within $\pm~10\degr$ of the image +Y axis (P.A. = 274\fdg9; arcs bounded in orange). B) 5$\arcsec$ x 5$\arcsec$ field, $r = 1\farcs4$ (red), $r = 2\arcsec$ (green), $r = 4\arcsec$ (yellow),  0 to 1 e$^{-}$~s$^{-1}$~pixel$^{-1}$ square root display.  C) 10$\arcsec$ x 10$\arcsec$ field, $r = 2\arcsec$ (red), $r = 4\arcsec$ (green), $r = 9\arcsec$ (yellow), [-3.3] to [+0.0] e$^{-}$~s$^{-1}$~pixel$^{-1}$ log$_{10}$ display.  Within $\pm$ 45$\degr$ of North and 4$\arcsec$-- 9$\arcsec$ from the star a
faint nebulosity can be seen, with some suggestion of curvature toward the east.}
\end{figure}

\clearpage
\begin{figure}
\epsscale{1.0}
\plotone{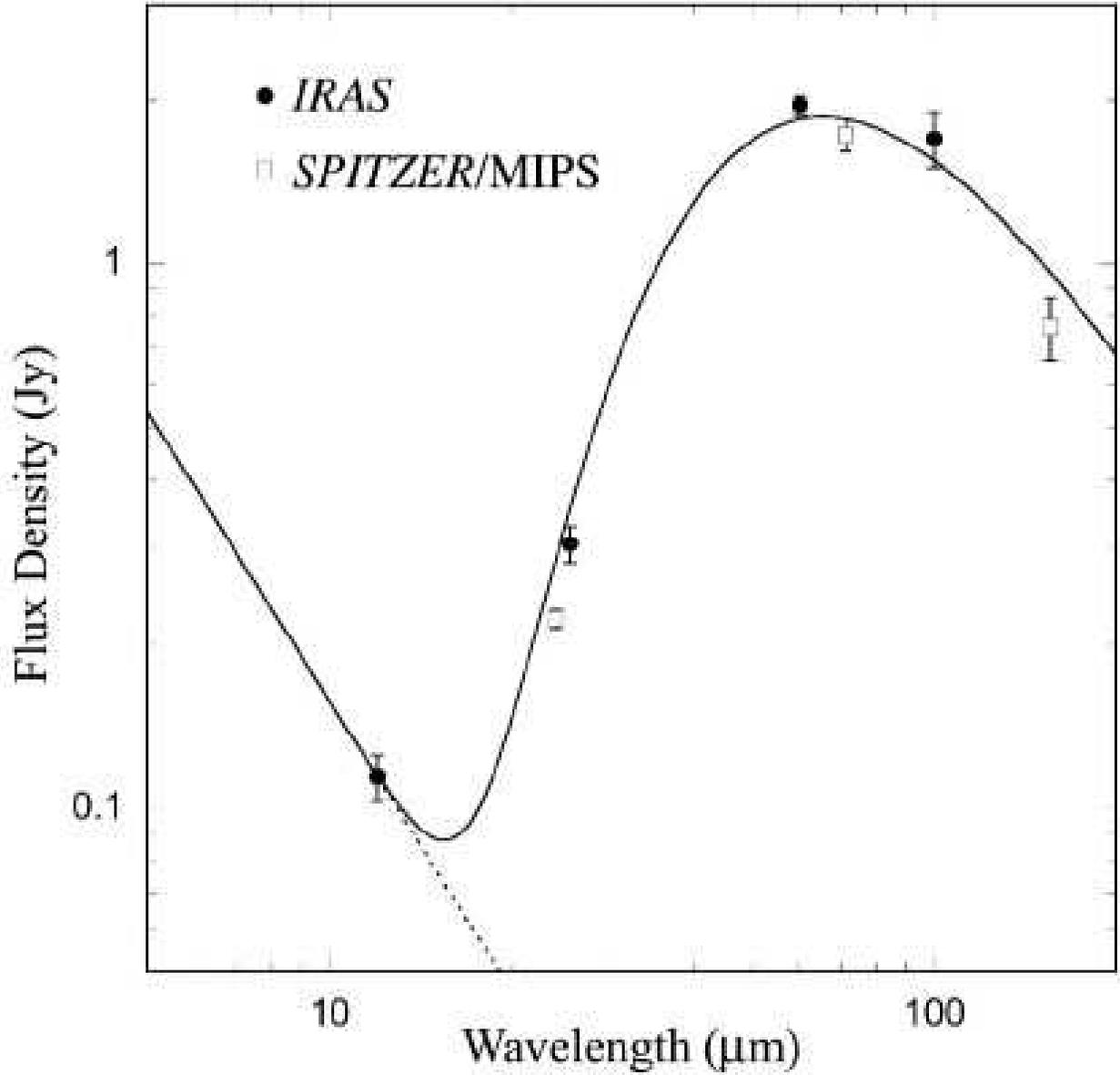}  
\caption{HD 181327 spectral energy distribution from color-corrected {\it Spitzer}/MIPS and {\it IRAS} photometry (see Table 2).  A least-squares blackbody fit to the photometric data indicates a grain equilibrium temperature of 77K $\pm$ 5K. The black line is the sum of the 77K fit and a blackbody of T$_{eff}$ = 6450K for the stellar photosphere (dotted line).}
\end{figure}

\clearpage
\begin{figure}
\epsscale{1.00}
\plotone{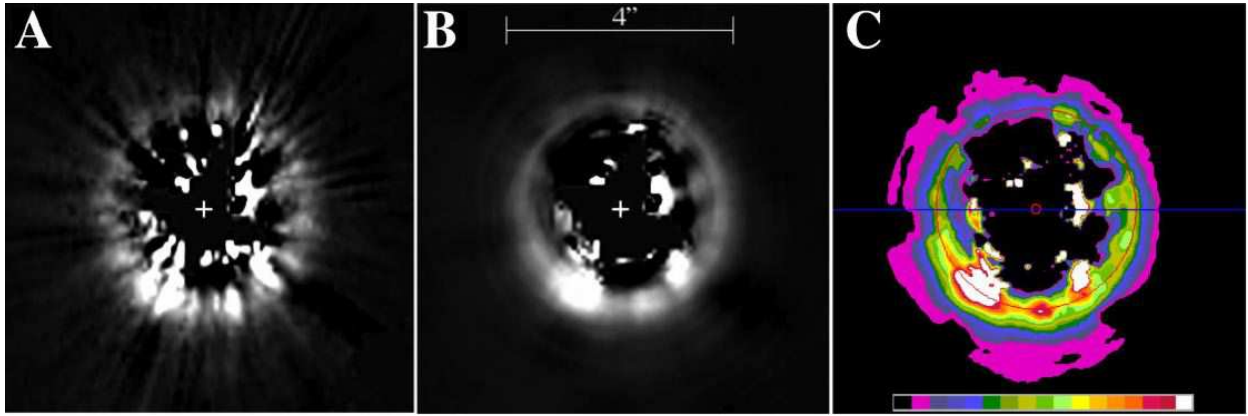}  
\caption{HD 181327 debris ring projected to a ``face-on'' geometry while conserving the total flux density. Panels (A) and (B) in this figure correspond to panel (A) and (C) in figure 2, but with the disk rotated to place the disk major axis (as seen on the sky) along the image horizontal before projection.  Panel C is a linear  isophotal contour map map of the image in panel B with 6.25\% SB contour intervals from +0.1 to to +2.19 mJy arcsec$^{-2}$.  Regions shown as hard white (SB $\geq$ 2.2 mJy arcsec$^{-2}$) are strongly affected by PSF-subtraction artifacts and should be ignored. The ``+'' symbols and small red circle mark the location of the star as determined from the target acquisition images. The large red circle corresponds to the distance at which the surface brightness peaks radially, and the blue line indicates the major axis of the ring as projected onto the sky.}
\end{figure}

\clearpage
\begin{figure}
\epsscale{1.00}
\plotone{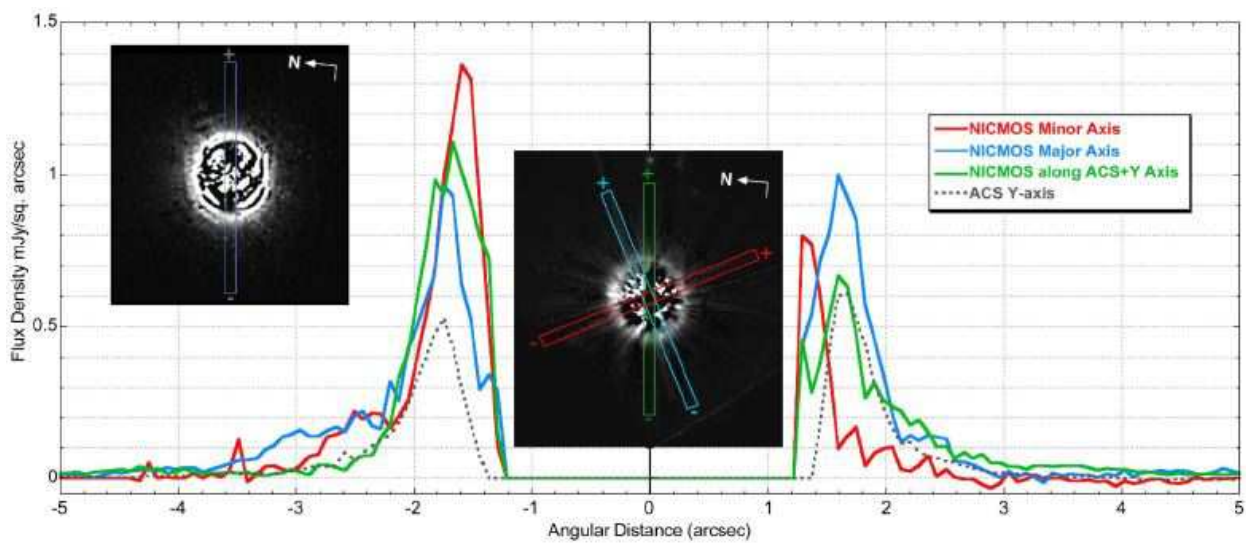}  
\caption{Radial surface brightness profiles of the HD 181327 circumstellar ring. NICMOS measures along the disk minor and major axes, and in the direction of the ACS image +Y axis,
in 0$\farcs5$ wide strips in radial increments of 1 pixel ($\sim$ 75.8 mas).  ACS measures made identically along the HRC image +Y axis (NICMOS and ACS images displayed at the same 
spatial scale and celestial orientation). Uncertainties in the NICMOS measures may be estimated from the 1 $\sigma$ errors shown in Figure 8. Accurately deriving the ACS 
measurement uncertainties is problematic due to the very strong azimuthal dependence of the residuals.  Based on the profile asymmetries we estimate the 
uncertainties in the ACS measures in the 1\farcs5 -- 2\farcs2 radial region to be  $<\pm$ 30\% of the SB.}
\end{figure}

\clearpage
\begin{figure}
\epsscale{1.00}
\plotone{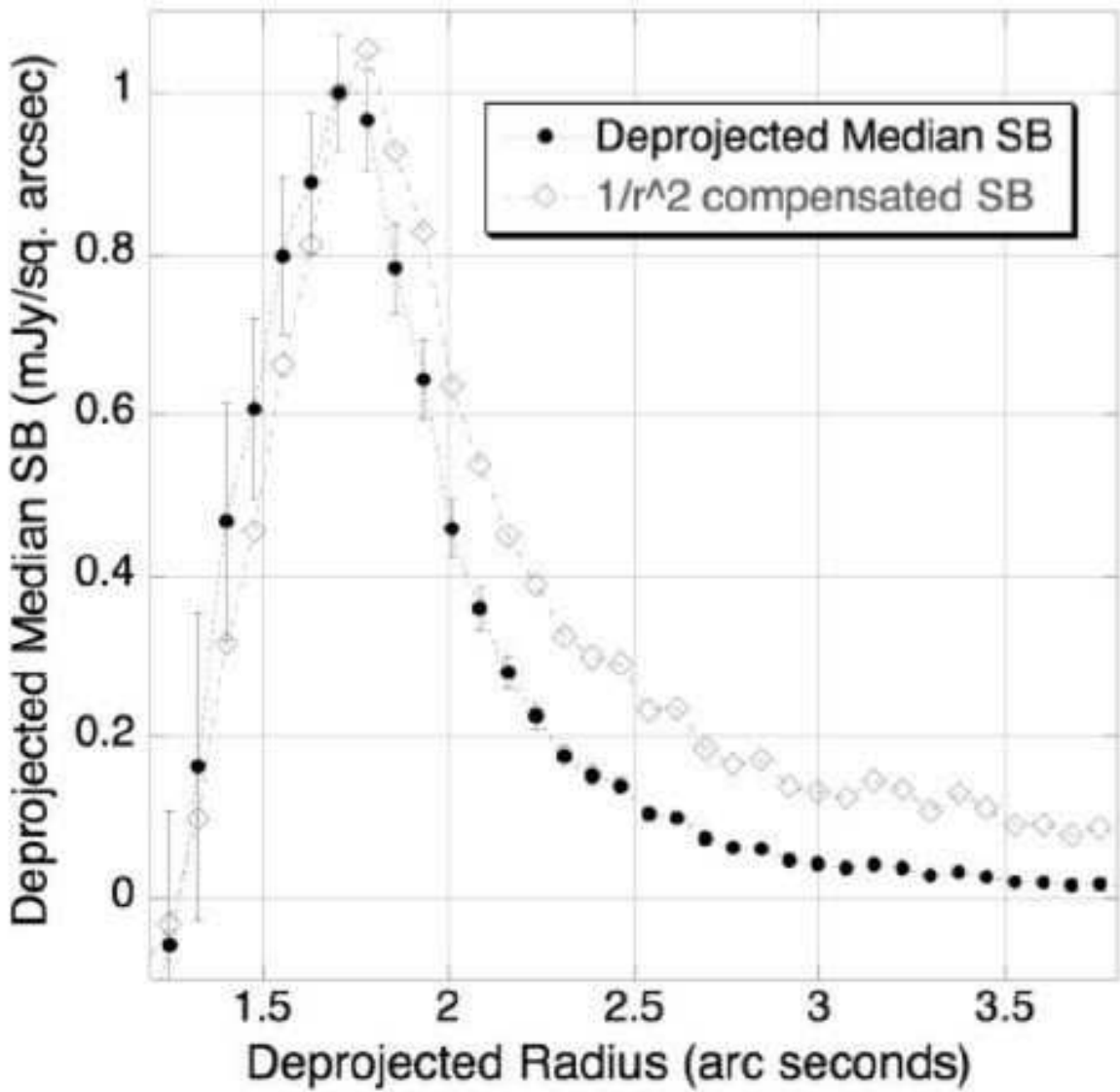}  
\caption{Azimuthally medianed 1.1 $\mu$m radial surface brightness profile of the HD 181327 ring (deprojected from an inclination angle of  31\fdg7 from face on) and the 1 $\sigma$  uncertainties in each 75.8 pixel wide annular zone measured. The 1 $\sigma$ error bars indicate the dispersion in the NICMOS surface brightness measures around the ring at each elliptical annulus, and does not include the $\sim$ 9\% uncertainty in the absolute calibration of the surface brightness. An adjusted profile compensating for the dilution of starlight with distance from the central star normalized to the radius of peak brightness is also shown.}
\end{figure}

\clearpage
\begin{figure}
\epsscale{1.00}
\plotone{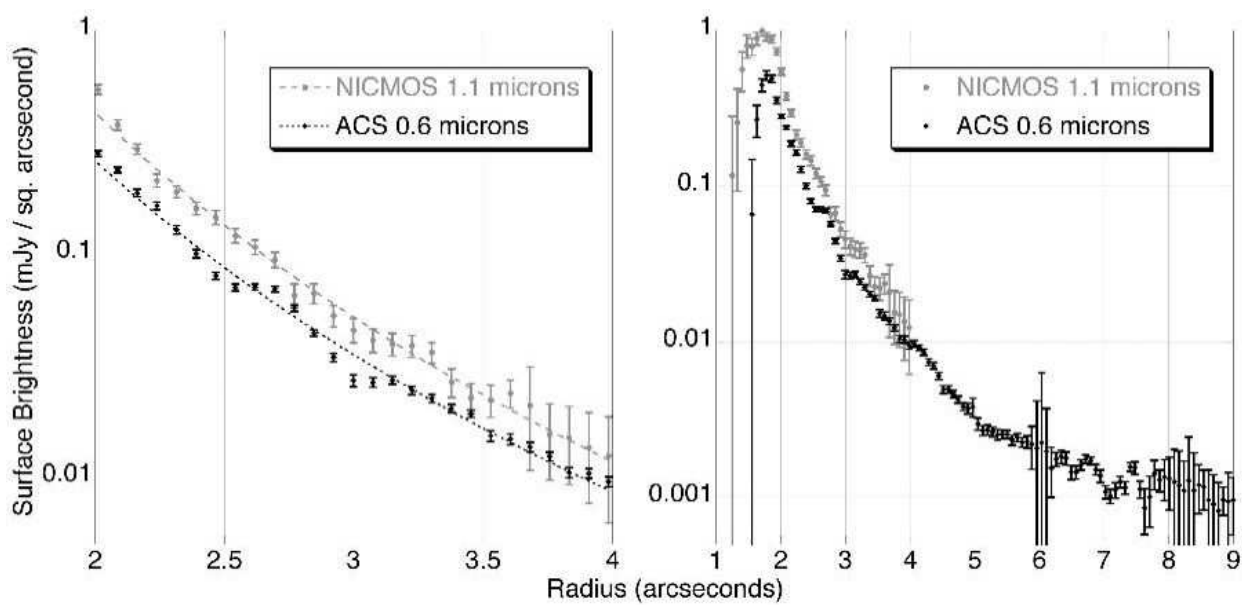}  
\caption{NICMOS and ACS deprojected SB profiles of the HD~181327 circumstellar disk from: Left --  2\arcsec\ to 4\arcsec~($\sim$ 100 to 200 AU), Right -- inner detection radii to 9\arcsec~($\sim$ 450 AU). NICMOS profile limited to {\it r} $<$ 4\arcsec~due to field of view truncation (see figure 9) and integration depth.   1 $\sigma$ error bars are the standard deviations of the measured SB in each 1 pixel wide radial zone about the zonal median and do not include uncertainties in the absolute calibration of the disk flux densities.  The ACS profile was measured after resampling the image to the NICMOS pixel scale (effectively smoothing the image with an $\approx$ 3x3 pixel kernel). }
\end{figure}

\clearpage
\begin{figure}
\epsscale{1.00}
\plotone{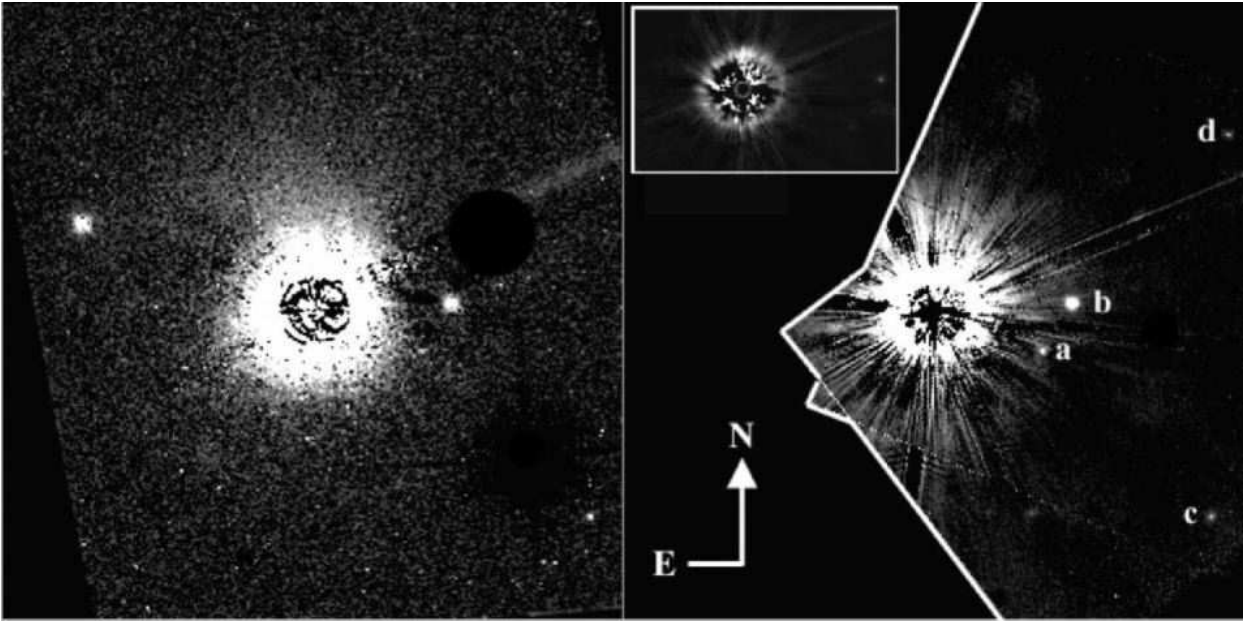}   
\caption{The outer regions of the HD~181327 disk in the ACS (left) and NICMOS (right) PSF-subtracted images within $\pm$ 12\farcs5 of the central star. Both images are square-root displays with maximum intensities (hard white) corresponding to 1.4 instrumental counts s$^{-1}$ pixel$^{-1}$. Four stars common in both fields are noted ({\it a--d}). The debris ring revealed in the NICMOS image is shown at the same spatial scale and image orientation in the inset (linear display from -0.25 to +10 counts s$^{-1}$ pixel$^{-1}$).}
\end{figure}

\clearpage
\begin{figure}
\epsscale{1.00}
\plotone{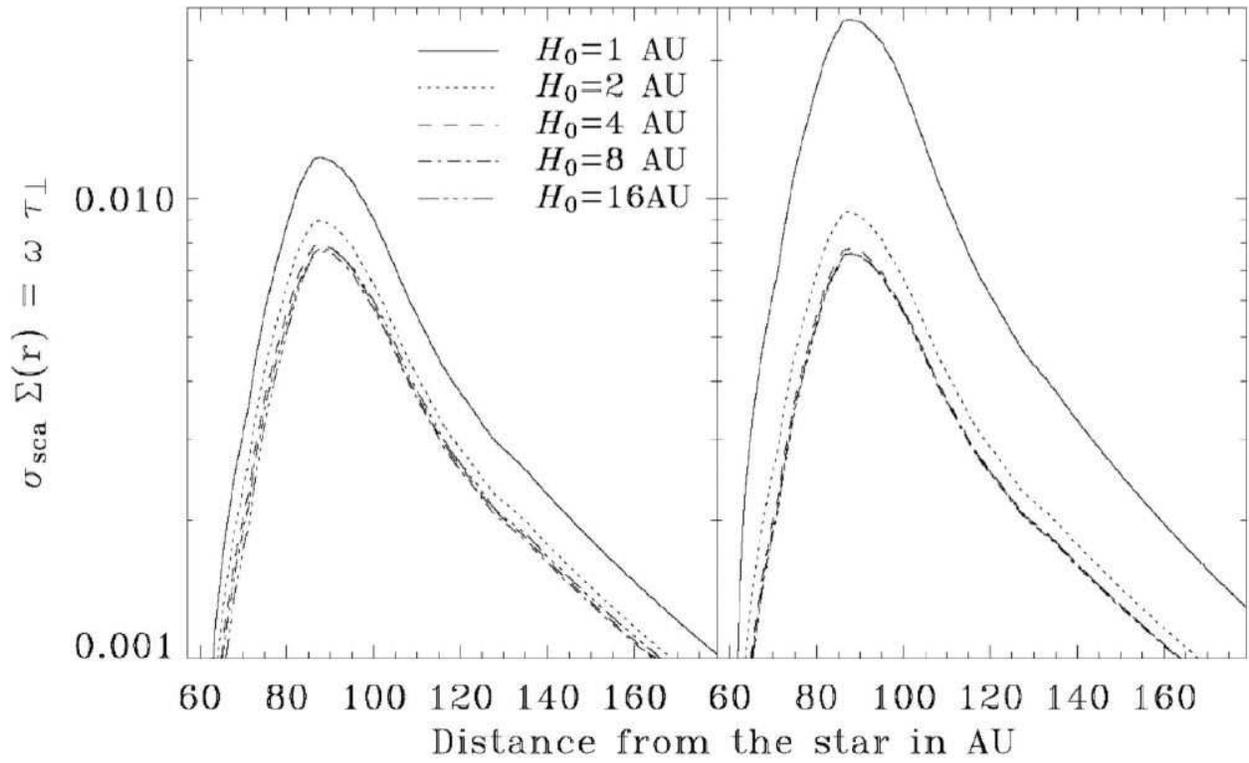}  
\caption{Surface density profiles multiplied by the mean scattering cross section at 1.1 $\mu$m assuming different vertical scale-heights ({\it H}$_{0}$). Left: exponential vertical profile ($\gamma$ = 1).  Right: Gaussian profile ($\gamma$ = 2).}
\end{figure}

\clearpage
\begin{figure}
\epsscale{1.00}
\plotone{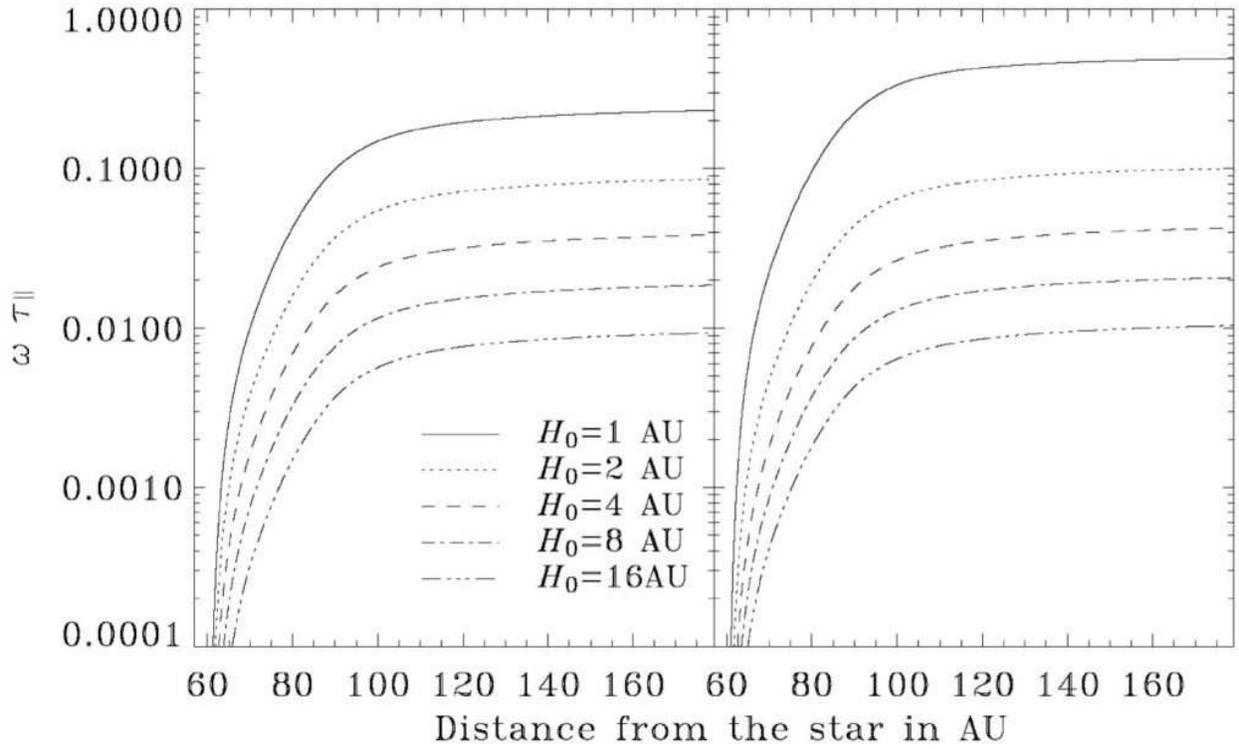}  
\caption{Midplane disk thickness in the NICMOS F110W band as a function of the distance from the star for different vertical scale-heights ({\it H}$_{0}$). Left: exponential vertical profile ($\gamma$ = 1).  Right: Gaussian profile ($\gamma$ = 2).}
\end{figure}

\clearpage
\begin{figure}
\epsscale{1.00}
\plotone{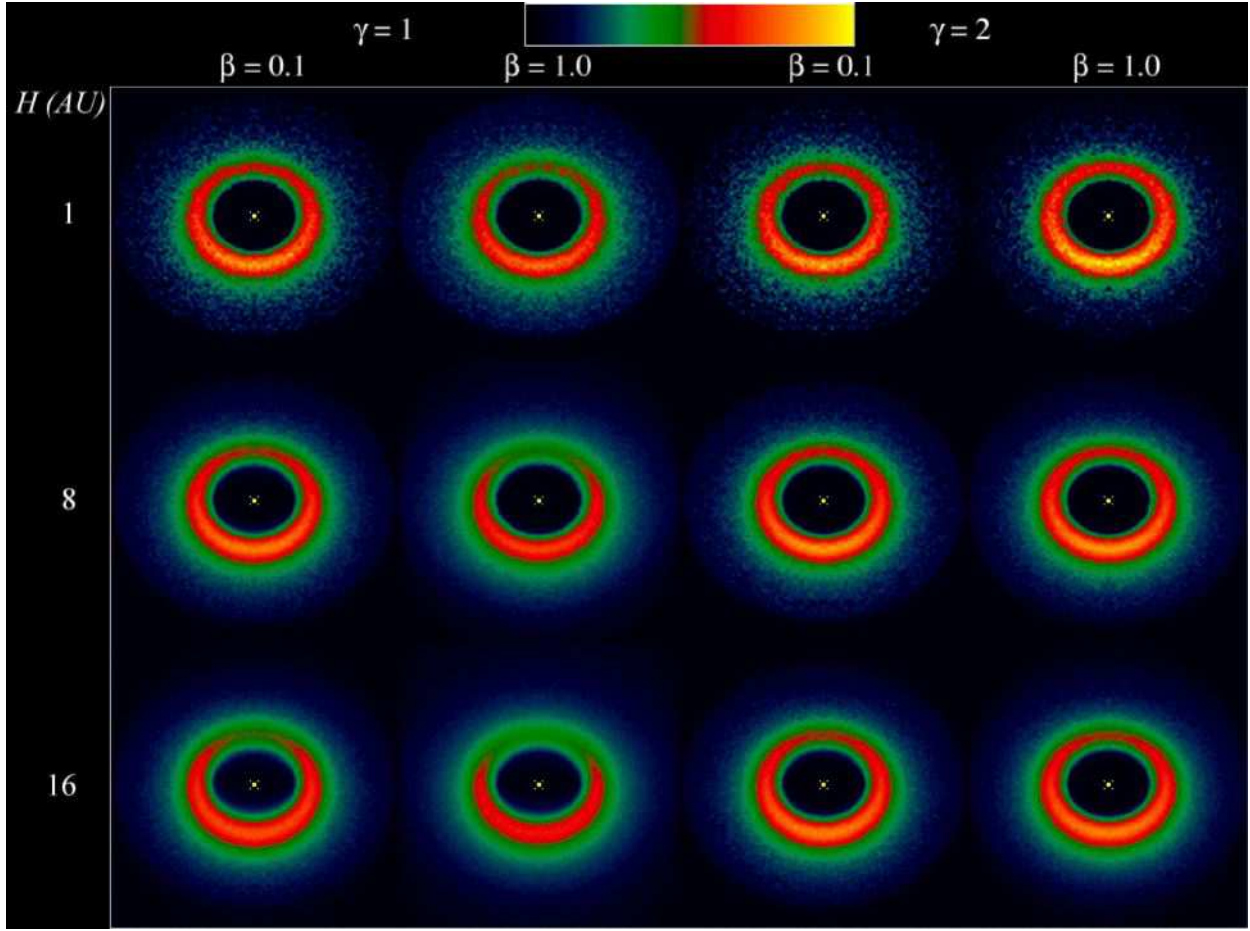}  
\caption{Scattered-light patterns from representative nearly-flat ($\beta$ = 0.1) and linearly flaring ($\beta$ = 
1) model disks with both exponential ($\gamma$ = 1) and Gaussian ($\gamma$ = 2) midplane 
number density profiles for disks with vertical  scale heights, $H_{0}$, of 1, 8, and 16 AU at 
$r_{0}$ ($r_{0}$ = 86.3 AU for HD~181327). Square-root stretched image displays normalized to 
the peak intensity of the ($\beta = 1, \gamma = 2, H_{0} = 1$AU) model disk (upper right panel).}
\end{figure}

\clearpage
\begin{figure}
\epsscale{1.00}
\plotone{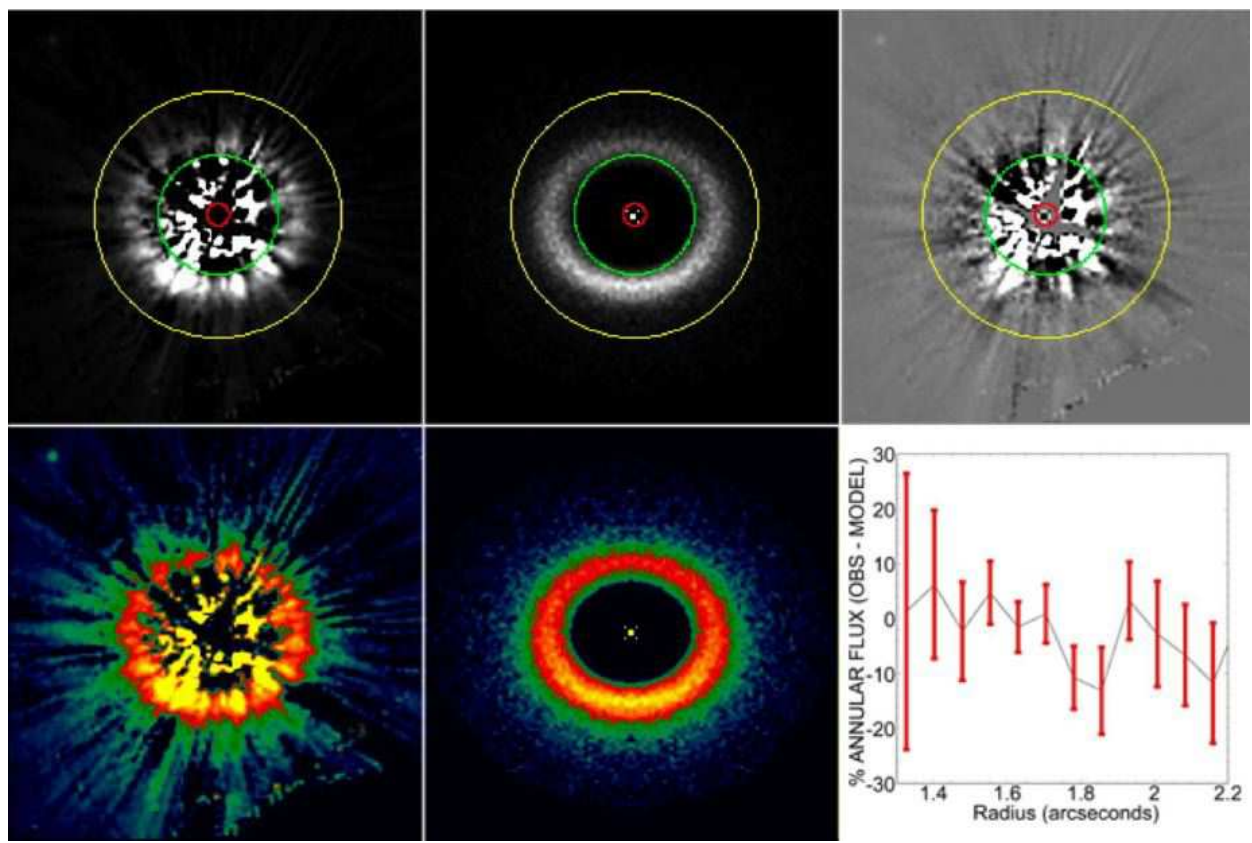}  
\caption{A  representative $\chi^{2}$ (convergent minimization) fit of one of the MCFOST  disk models (middle panels;  $\beta = 1, \gamma = 2, H_{0} = 1$ AU case as shown in Fig.12)  to the NICMOS observation of HD~181327 (left panels).  The fitting region (in the $r \approx 1\farcs3 - 2\farcs3$ annular zone) is bounded by the green and yellow circles overlaid in the  upper panels.  The upper image panels are linear displays (0 to 1.5 mJy arcsec$^{-2}$, black to white,  closely corresponding to the peak surface brightness of the HD 181327 disk. The lower image panels show the same data displayed over the same dynamic range but with a square-root stretch (same color-intensity scale as shown in Figure 12).  The residuals remaining after subtracting the  model image are shown in the linear  ($\pm$1.5 mJy arcsec$^{2}$) display in the upper right panel.  The residual flux densities and their 1 $\sigma$ deviations about their respective medians in each one pixel (75.8 mas) wide annular zone, as a percentage of the predicted flux densities ($\sim$ 10\% over the region of the scattered light ring), are shown in the lower right panel.}
\end{figure}

\clearpage
\begin{figure}
\epsscale{1.00}
\plotone{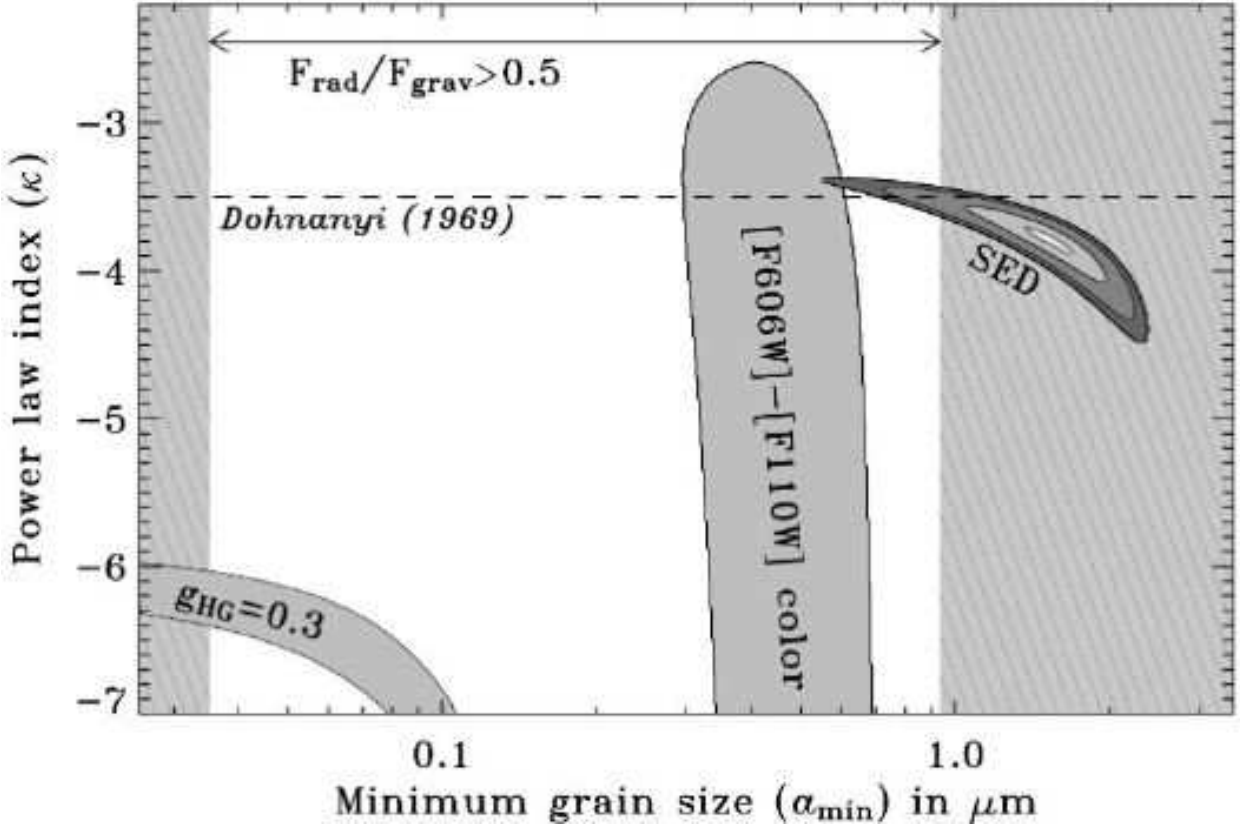}  
\caption{The minimum grain sizes and power law indices suggested, separately, from the observed scattered light disk asymmetry factor ($\S$ 5.3), the [0.6] - [1.1] $\mu$m color index of the grains ($\S$ 6.1), the thermal IR spectral energy distribution ($\S$ 4), and a theoretical radiation force blow-out size limit ($\S$ 8) presuming a single population of hard spherical silicate grains with d{\it n}({\it a}) $\propto$ {\it a}$^{\kappa}$~d{\it a}. Optical constants from  \cite{Wein01} are adopted for Mie scattering by the grains.  Radiation force and grain temperatures have been calculated with a NextGen atmosphere model spectrum ({\it T}$_{eff}$ = 6450 K, log($g$) = 4.5; \cite{Hau99}) scaled to match the observed V-band magnitude. SED fits corresponding to reduced  $\chi^{2}$ of 0.25, 1, 2.5, and 3.5 (with corresponding probabilities of 78\%, 37\%, 8\%, and 3\%) are shown.}
\end{figure}

\clearpage
\begin{figure}
\epsscale{.85}
\plotone{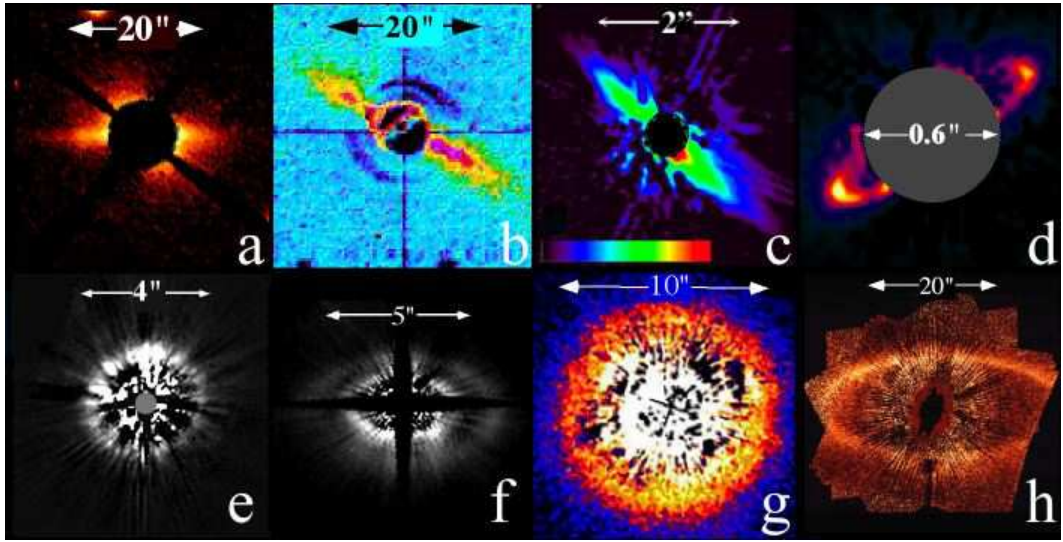}  
\caption{Circumstellar debris systems imaged in scattered light. (a) AU Mic, (b) $\beta$ Pic, (c) HD~32297( d) HR~4796A, (e) HD181327, (f)
HD~141569, (g) HD~107146, (h) $\alpha$ PsA. See Table 3 for details.}
\end{figure} 


\begin{thebibliography}{}

\bibitem[Ardila et al.(2004)]{ardila04} Ardila, D. R., et al. 2004, \apj, 617, L147
\bibitem[Artymowicz \& Lubow(1994)]{arty94} Artymowicz, P., \& Lubow, S. H. 1994, \apj, 428, 581
\bibitem[Augereau et al. (1999)]{Aug99}Augereau, J. C., Lagrange, A. M., Mouillet, D., Papaloizou,
J. C. B., \& Grorod, P. A. 1999, A\&A, 348, 557
\bibitem[Augereau et al. (1999b)]{Aug99b}Augereau, J. C., Lagrange, A. M., Mouillet, D., Menard, F. 1999, A\&A, 350, L51
\bibitem[Augereau \& Beust (2006)]{Aug06}Augereau, J. C. \& Beust, H. 2006 A\&A, in press (astro-ph/0604313)
\bibitem[Augereau et al.(2001)]{aug01}Augereau, J. C.. Nelson, R. P., Lagrange, A. M., Papaloizou, J. C. B., Mouillet, D. 2001, A\&A, 370, 447
\bibitem[Aumann et al.(1984)]{aumann84}Aumann, H. H., Beichman, C. A, Gillett, F. C., de Jong, T., Houck, J. R., Low, F. J., Neugebauer, G, Walker, R. G., \& Wesselius, P. R. 1983, \apjl, 278, 23
\bibitem[Backman \& Paresce (1993)]{BAC93}Backman, D. E., \& Paresce, F. 1993, 
in Protostars and Planets III, ed. E. H. Levy \& J. I. Lunine (Tucson: Univ. Arizona Press), 1253
\bibitem[Baraffe et al.(1998)]{ba98}Baraffe, I., Chabrier, G., Allard, F., \& Hauschildt, P. 1998, A\&A, 337 403
\bibitem[Bohlin et al.(2003)]{bohlin02}Bohlin, R., Hartig, G., \& Sparks, W. 2003, STScI Instument Science Report ACS 02-03, http://www.stsci.edu/hst/acs/documents/isrs/isr0203.pdf 
\bibitem[Buitrago \& Mediavilla(1986)]{buit}Buitrago, J., \& Mediavilla, E. 1986, A\&A, 169, 95
\bibitem[Burgasser et al.(2006)]{burgasser05} Burgasser, A. J., Reid, I. N., Siegler, S.,  Close, L., Allen, P., Lowrance, P., and Gizis J. 2006, in Protostars \& Planets V, ed. B. Reipurth, D. Jewitt, \& K. Keil (Tucson: Univ. of Arizona Press), in press.
\bibitem[Chauvin et al.(2003)]{CHA04}Chauvin, G., et al. 2003, A\&A, 404, 157
\bibitem[Chen et al.(2006)]{chen}Chen, C. 2006, ApJ, in press (astro-ph/0605277)
\bibitem[Clampin et al.(2003)]{CLA03}Clampin, M., et al. 2003, AJ, 126, 385 
\bibitem[Close et al.(2003)]{close03}Close, L. M.,Siegler, N.,Freed, M.,Biller, B. 2003, \apj, 587, 407
\bibitem[Cutri et al.(2003)]{2MASS}Cutri, R.~M., et al.\ 
2003, VizieR Online Data Catalog, 2246, 0 
\bibitem[D'Antona \& Mazzitelli(1997)]{dm97} D'Antona, F. \& Mazzitelli, I. 1997, MmSAI, 68, 807
\bibitem[Decin et al.(2003)]{DEC03}Decin, G., Dominik, C., Waters, L. B. F. M., \& Waelkens, C. 2003, \apj, 598, 636 
\bibitem[Dohnanyi(1969)]{Doh69}Dohnanyi, J. W. 1969, J. Geophys. Res., 74, 2531
\bibitem[Gilliland(2004)]{gill04}Gilliland, R. 2004,  STScI Instument Science Report ACS 04-01, http://www.stsci.edu/hst/acs/documents/isrs/isr0401.pdf
\bibitem[Gonzaga et al.(2005)]{gonz}Gonzaga, S., et al. 2005, ``ACS Instrument Handbook'', 6.0, (Baltimore:STScI)
\bibitem[Grady et al.(2001)]{grady2001}Grady, C. A., et al. 2001, \aj, 122, 3396
\bibitem[Hauschildt et al. (1999)]{Hau99}Hauschildt, P. H., Allard, F., \& Baron, E. 1999, \apj, 512, 377
\bibitem[Henyey \& Greenstein(1941)]{HG41}Henyey L. G., Greenstein J. L. 1941, \apj, 93, 70
\bibitem[Hines \& Schneider(2004)]{Hines04}Hines, D. C., \& Schneider, G. 2004, AAS, 205.0504
\bibitem[Kalas, et al.(2000)]{KAL00}Kalas, P., Larwood, J., Smith, B. A., \& Schultz, A. 2000, \apj, 530, L133
\bibitem[Kalas, et al.(2002)]{KAL02}Kalas, P., Graham, J. R., Beckwith, S. V. W., Jewitt, D. C. \& Lloyd, J. P., 2002, \apj, 567, 999
\bibitem[Kalas, et al.(2004)]{KAL04}Kalas, P., Liu, M. C., Matthews, B. C. 2004, Science, 5666, 1990
\bibitem[Kalas, et al.(2005)]{KAL05}Kalas, P., Graham, J.R., \& Clampin, M. 2005, Nature, 435, 1067
\bibitem[Kalas, et al.(2006)]{KAL06}Kalas, P., et al. 2006, ApJ, 637, L60
\bibitem[Kalas \& Jewitt (1995)]{kal95}Kalas, P. \& Jewitt, D. 1995, \aj, 110, 794
\bibitem[Klahr \& Lin (2005)]{klar}Klahr, H., \& Lin, D. N. C. 2005,  ApJ, 632,111
\bibitem[Krist \& Hook (1997)]{HOOK97}Krist, J., \& Hook, R. N. 1997, in The 1997 Calibration Workshop, eds. S. Casertano, R. Jedrzejewski, and T. Keyes (Baltimore: STScI), 192
\bibitem[Krist et al.(2005)]{krist05}Krist, J., et al. 2005, AJ, 129, 1008
\bibitem[Lecavelier des \'Etangs et al.\ (1996)]{Lec96}Lecavelier des \'Etangs, A., Vidal-Madjar, A., \& Ferlet, R. 1996, A\&A, 307, 542
\bibitem[Liu(2004)]{liu04}Liu, M. 2004, Science, 305, 1442
\bibitem[Lowrance et al.(2000)]{Lowrance2000}Lowrance, P. J., Schneider, G., Kirkpatrick, J. D., Becklin, E. E., Weinberger, A. J., Zuckeran, B., Plait, P., Malmuth, E. M., Heap, S. R., Schultz, A., Smith, B. A.,Terrile, R. J., Hines, D. C. 2000 \apjl, 541, 390
\bibitem[Lowrance et al.(2005)]{Lowr05}Lowrance, P. J., et al., 2005, \aj, 130, 1845
\bibitem[Mannings \& Barlow\ (1998)]{MAN98}Mannings, V., \& Barlow, M. J. 1998, ApJ, 497, 330
\bibitem[Mathis, Rumpl \& Nordsieck (1977)]{mathis}Mathis, J., Rumpl, W., \& Nordsieck, K. 1977, ApJ, 217, 425
\bibitem[Mamajek et al.\ (2004)]{MAM04}Mamajek, E., et al. 2004, ApJ, 612, 496
\bibitem[Marsh et al.(2002)]{Marsh}Marsh, K. A., Silverstone, M. D., Becklin, E. E., Koerner, D. W.;, Werner, M. W., Weinberger, A. J.,  Ressler, M. E. 2002,  \apj, 573,425
\bibitem[Metchev et al.(2005)]{metch05}Metchev, S. A. et al. 2005, \apj, 622, 451
\bibitem[Mouillet et al.(2001)]{mou01}Mouillet, D., Lagrange, A. M., Augereau, J. C., M\'enard, F. 2001, A\&A, 327, 61
\bibitem[Neuh{\"a}user et al.\ (2003)]{NEU04}Neuh{\"a}user, et al. 2003, Astron. Nachr. 324,  535
\bibitem[Noll, K. et al.\ (2004)]{NOLL04}Noll, K., et al. 2004, "NICMOS Instrument Handbook", Version 7.0, (Baltimore: STScI) 
\bibitem[Nordstrom et al.\ (2004)]{nord04}Nordstrom, B.,  Mayor, M., Andersen, J.,Holmberg, J.,Pont, F., Jorgensen, B. R., Olsen, E. H., Udry, S., \& Mowlavi, N., et al. 2004, A\&A. 418, 989
\bibitem[Ozernoy et al. (2000)]{ozer}Ozernoy, L. M., Gorkavyi, N. N., Mather, J. C., \& Taidakova, T. A. 2000, ApJ, 537, L147
\bibitem[Palla \& Stahler(2001)]{ps01}Palla, F. \& Stahler, S. W. 2001, \apj, 533, 299
\bibitem[Pantin \& Lagage(1996)]{pantin96}Pantin, E., \& Lagage, P.-O. 1996, in The Role of Dust in the Formation of Stars, eds. H. U.~K{\"a}ufl and R. Siebenmorgen (Berlin: Springer-Verlag)
\bibitem[Perryman et al.(1997)]{peri97}Perryman, M.A.C., et al. 1997, A\&A, 323,49
\bibitem[Pinte et al.(2006)]{pinte06}Pinte, C., et el. 2006, A\&A, submitted
\bibitem[Schneider et al.(1999)]{schn4796}Schneider, G., et al. 1999,  \apjl, 513,127
\bibitem[Schneider et al.(2001)]{schn01}Schneider, G., et al. 1999,  \aj, 121,525
\bibitem[Schneider et al.\ (2004)]{schgil}Schneider, G., Cotera, A. S., Silverstone, M. D., \& Weinberger, A. J. 2004, ASP Conf. Ser., 324, 247
\bibitem[Schneider et al.(2003)]{SSH03} Schneider, G., Silverstone, M. D., and Hines, D. C. 2003,  BAAS, 2003.4602S
\bibitem[Schneider (2004)]{sch04}Schneider, G. 2004, ASP Conf. Ser 196 242
\bibitem[Schneider et al.(2005)]{SCH05} Schneider, G., Silverstone, M. D., and Hines, D. C. 2005,  \apjl, 629, L117
\bibitem[Schneider et al.(2005b)]{SCH05B} Schneider, G., Cotera, A S., Silverstone, M. D., Weinberger, A. J., 2005b,  ASP Conf. Series, 324, 24.
\bibitem[Schneider \& Silverstone (2003)]{Schneider03}Schneider, G., \& Silverstone, M. 2003,  in High-Contrast Imaging for Exo-Planet Detection, eds. A. B. Schultz \& R. G. Lyon, SPIE Proceedings Series, 4860, 1.
\bibitem[Smith \& Terrile (1984)]{SMI84}Smith, B. A., \& Terrile, R. J. 1984, Science, 226, 1421
\bibitem[Takeuchi \& Artymowicz(2001)]{tag}Takeuchi, T, \& Artymowicz, P, 2001, ApJ, 557, 990
\bibitem[Th\'ebault \& Augereau (2005)]{theb05}Th\'ebault, P., \& Augereau, J-C. 2005, A\&A, 437, 141
\bibitem[Telesco et al. (2000)]{telesco}Telesco, C., et al. 2000, ApJ, 530, 329
\bibitem[Weinberger et al.(1999)]{Wein99}Weinberger, A., et al. 1999, \apj, 525, L53
\bibitem[Weingartner \& Draine (2001)]{Wein01}Weingartner, J. C., \& Draine, B. T. 2001, \apj, 548, 296
\bibitem[Zuckerman \& Webb (2000)]{ZUK00}Zuckerman, B., \& Webb, R. A. 2000, ApJ, 535, 959
\bibitem[Zuckerman et al.(2001)]{ZUK01}Zuckerman B., \& Song, I., Bessell, M. S., \& Webb, R. A., 2001, ApJ, 562, 87
\bibitem[Zuckerman \& Song (2004)]{ZUK04}Zuckerman B., \& Song, I. 2004, ApJ, 603, 738

\end{thebibliography}
\end{document}